\documentclass[12pt, draftclsnofoot, one column]{IEEEtran}

\usepackage[cmex10]{amsmath}
\usepackage{amsfonts,amssymb,cite,graphicx,delarray,color}
\usepackage{multirow}
\usepackage{algorithmicx}
\usepackage{algorithm}
\usepackage{algpseudocode}
\usepackage{array}
\usepackage{float}
\usepackage{subfigure}
\usepackage{units}
\usepackage{mathrsfs}
\usepackage{xcolor}

\IEEEoverridecommandlockouts

\begin{document}

\title{Layered Group Sparse Beamforming for Cache-Enabled \textcolor{black}{Green} Wireless Networks}

\author{\IEEEauthorblockN{Xi~Peng,~\IEEEmembership{Student Member,~IEEE}, Yuanming~Shi,~\IEEEmembership{Member,~IEEE}, Jun~Zhang,~\IEEEmembership{Senior Member,~IEEE}, and Khaled~B.~Letaief,~\IEEEmembership{Fellow,~IEEE}} \thanks{X. Peng, J. Zhang and K. B. Letaief are with the Dept. of ECE at the Hong Kong University of Science and Technology, Hong Kong (email: \{xpengab, eejzhang, eekhaled\}@ust.hk). K. B. Letaief is also affiliated with Hamad bin Khalifa University, Doha, Qatar (e-mail: kletaief@hbku.edu.qa). Y. Shi is with the School of Information Science and Technology, ShanghaiTech University, Shanghai, China (e-mail: shiym@shanghaitech.edu.cn).} \thanks{This work is supported by the Hong Kong Research Grant Council under Grant No. 16200214.}}

\begin{titlepage}
\maketitle
\thispagestyle{empty}
\vspace{-1.5cm}
\begin{abstract}
The exponential growth of mobile data traffic is driving the deployment of dense wireless networks, which will not only impose heavy backhaul burdens, but also generate considerable power consumption. Introducing caches to the wireless network edge is a potential and cost-effective solution to address these challenges. In this paper, we will investigate the problem of minimizing the network power consumption of cache-enabled wireless networks, consisting of the base station (BS) and backhaul power consumption. The objective is to develop efficient algorithms that unify adaptive BS selection, backhaul content assignment and multicast beamforming, while taking account of user QoS requirements and backhaul capacity limitations. To address the NP-hardness of the network power minimization problem, we first propose a generalized layered group sparse beamforming (LGSBF) modeling framework, which helps to reveal the layered sparsity structure in the beamformers. By adopting the reweighted $\left.\ell_{1}\right/\ell_{2}$-norm technique, we further develop a convex approximation procedure for the LGSBF problem, followed by a three-stage iterative LGSBF framework to induce the desired sparsity structure in the beamformers. Simulation results validate the effectiveness of the proposed algorithm in reducing the network power consumption, and demonstrate that caching plays a more significant role in networks with higher user densities and less power-efficient backhaul links.
\end{abstract}

\begin{IEEEkeywords}
Wireless caching, content-centric wireless networks, multicasting beamforming, layered group sparse beamforming, convex approximation, network power minimization, green communications.
\end{IEEEkeywords}

\end{titlepage}
\newpage

\section{Introduction}

To cater for the unprecedented explosion of mobile data traffic \cite{Cisco}, cell densification has been regarded as a key mechanism for further wireless evolution \cite{netDensification_5G}. To effectively manage co-channel interference in dense cellular networks, coordinated multipoint (CoMP) technology, i.e., cooperation among base stations (BSs), has been proposed \cite{Lee2012_Coordinated_multipoint_LTE_A}. However, it requires data sharing among cooperative BSs, which will yield considerable backhaul traffic. Current small cell backhaul solutions, such as xDSL \cite{DSL_value} and non-line-of-sight microwave \cite{5Gwireless_bh}, are far from adequate to provide sufficient data rate and thus make the current networks vulnerable to congestion. Caching frequently requested content at the wireless network edge, especially at small BSs \cite{FemtoCaching2013}, has been recently proposed as a cost-effective approach to lower the latency for content delivery and alleviate the heavy burden on backhaul links. Remarkably, caching also has the prominent advantage in improving the network energy efficiency. Since local caching brings the content closer to mobile users (MUs) and enables content delivery without using backhaul links, BS transmit power and backhaul power can be substantially reduced. 

Energy efficiency, as an essential concern in green cellular networks, has attracted global attention \cite{GreenMobileNetSurvey} since it is related to maintaining profitability for cellular operators, as well as reducing the overall environmental effects. Most previous investigations on energy efficiency of cellular networks either ignored the backhaul power consumption \cite{Richter2010vtc} or employed simplified models to measure it \cite{ZhuangFuxing2014}. As cellular networks will evolve to be progressively dense and heterogeneous, backhaul power consumption will play an increasingly important role in total network power consumption \cite{BHpower}. It is inspiring that caching can be very effective in fundamentally reducing backhaul power consumption. Owing to the recent technology development of caching hardware \cite{CacheEnergyEfficiency}, massive backhaul data can be reduced with energy-efficient caches. \textcolor{black}{Also, the frequent reuse of cached contents implies the potential of cache-enabled networks in energy saving.} In this paper, we will investigate network power minimization for cache-enabled wireless networks, by taking both the BS and backhaul power consumption into consideration.

\subsection{Related works}
Caching popular contents at small BSs has been attracting a lot of attention. The idea of femtocaching was first proposed in \cite{FemtoCaching2012} to alleviate backhaul loads for small BSs with low-capacity backhaul links. The caching content could be uncoded or coded, and a coded caching scheme can achieve a global caching gain as discussed in \cite{FemtoCaching2013}. 
However, these initial studies \textcolor{black}{assumed no interference among different communication links, and did not take the impact of wireless channels into account. It was proposed in \cite{Maddah2015ISIT_CacheAided,Maddah2017Cache-Aided} that caching at BSs will not only provide load balancing gain, but also bring interference cancellation gain and interference alignment gain. Follow-up papers \cite{Li2015distributed,Juan2016ICC_distributedCaching,bacstug2015cache} have shed light on cache-aided wireless communications and interference management under various performance metrics. Aiming at minimizing the download delay, distributed caching algorithms were designed in \cite{Li2015distributed,Juan2016ICC_distributedCaching}.  The tradeoff between the small BS density and total cache size under a certain outage probability was investigated in \cite{bacstug2015cache}. 
Cooperation among multi-antenna cache-enabled BSs \cite{mixed-Precoding-Cache,Tao2015arXiv,SenguptaTS2017_FogAided} is promising since caching can reduce the backhaul requirement. Full cooperation was considered in \cite{mixed-Precoding-Cache} to minimize total transmit power.  
Dynamic clustering and partial cooperation were adopted in \cite{Tao2015arXiv}.  Moreover, by employing the cloud processing and edge caching, cooperative transmission and low delivery latency can be achieved at the same time \cite{SenguptaTS2017_FogAided}.}

There is a growing concern on energy efficiency in wireless networks. Previous works include transmit power minimization via coordinated beamforming \cite{MIMO_clusterLinearPrecoding,MIMO_limitedCoordination,HongLuoZQ2013,SparseBF_userCentricCluster} and adaptive selection of active BSs \cite{BS_PowerModel,GreenMobileNetSurvey,sleepmode,Wu2013Traffic-Aware}.
After introducing edge caches, similar approaches have been extended to the cache-enabled wireless networks \cite{mixed-Precoding-Cache,Yao2015novel}. 
With cell densification, backhaul power consumption will become a significant component of the total network power consumption \cite{Backhaul_cost}. \textcolor{black}{In \cite{Atzeni2017_flexibleCacheAided}, energy efficiency for cache-aided networks was optimized by assuming constant transmit power for small cell BSs and wireless backhaul nodes.}  In \cite{Poularakis2015_ExploitingCachingMulticast}, caching content placement and multicast association were optimized in order to minimize the overall energy cost. But it only considered
the backhaul power of the macro BS and did not count small BSs. In order to minimize the network power consumption, joint beamforming and backhaul data assignment problem was investigated in \cite{SparseBF_userCentricCluster,ZhuangFuxing2014,MyPIMRC,GSBF,smoothedLp,Tao2015arXiv}. But a comprehensive consideration of traffic-dependent backhaul power consumption, active BS selection, multicast beamforming, and backhaul data assignment is still missing.

There are some preliminary studies on developing sparsity-based approaches for designing wireless networks. Inspired by the success of sparse signal processing techniques such as compressed sensing \cite{CS_Donoho,CS_CandesTao}, more structured sparsity patterns have been exploited, including group sparsity \cite{GroupLasso}, overlapping group sparsity \cite{Overlap_GroupLasso}, and layered group sparsity \cite{Two-LayerSG,Multi-layer-GS-coding}, which yield efficient algorithms. Recent years have witnessed an increasing prevalence of applying sparse optimization to design wireless networks, such as the individual sparsity-inducing norm applied for user admission in \cite{smoothedLp} \textcolor{black}{and link admission control in \cite{Liu2015_JointPowerAndAdmission}}, and the group sparsity-inducing norm applied for active remote radio head selection of Cloud-RAN in \cite{GSBF}. Sparse optimization is further applied to joint beamforming and backhaul data assignment design in caching networks \cite{MyPIMRC,Tao2015arXiv}, which may provide potential solutions for 5G wireless networks. As will be revealed in this paper, network energy minimization in cache-enabled wireless networks involves more complicated sparsity structures, and thus more thorough investigations will be needed.

\subsection{Contributions}

The main objective of this work is to minimize the network power consumption
for cache-enabled wireless networks, which mainly consists of the
BS and backhaul power consumption. In this problem, coupled with the
non-convex combinatorial composite objective function, there are non-convex quadratic
QoS constraints due to multicast transmission, as well as the challenging
$\ell_{0}$-norm per-BS backhaul capacity constraints. As a result,
it is a mixed-integer non-linear programming problem, and is NP-hard. In this paper, we propose a systematic  framework to develop low-complexity
algorithms to solve this challenging problem. Specifically, our main contributions
are listed as follows:
\begin{enumerate}
\item We adopt a realistic model to evaluate the total network power consumption,
incorporating practical power consumption models for BSs and backhaul links. In
particular, we allow the BS sleep mode, and consider a traffic-dependent
backhaul power consumption model, which is essential to investigate backhaul-limited networks. To make the network power minimization problem tractable,
we propose a layered group sparse beamforming (LGSBF) modeling framework,
which is able to jointly select active BSs, assign backhaul data,
and determine the multicast beamformers. This generalized structured
sparse formulation unifies existing approaches \cite{MIMO_clusterLinearPrecoding,HongLuoZQ2013,GSBF,Tao2015arXiv}, and will assist the problem analysis and efficient algorithm design.
\item The LGSBF formulation reveals that adaptive BS selection (i.e., the decision for the active
BS set) and backhaul assignment (i.e., the delivery of uncached content
via backhaul links) can be achieved by controlling the sparsity structure
in multicast beamformers. To solve the problem, we first propose to convexify the original problem via structured group sparsity-inducing norm minimization. The second algorithmic contribution is an iterative search procedure that can effectively determine BS selection and backhaul assignment. Finally, coordinated multicast beamforming is adopted to determine the overall beamformers.
\item Simulation results are provided to demonstrate the effectiveness of
our proposed algorithm, and show the performance gain compared with
existing approaches, including the coordinated beamforming algorithm
\cite{CB} and two sparse multicast beamforming algorithms \cite{smoothedLp,Tao2015arXiv}.
Moreover, we observe that the network performance can be effectively
enhanced by employing edge caching, which shows the potential of caches
as effective and efficient alternatives for high-capacity backhaul
links. In particular, it is shown that caching can reduce the network
power consumption more effectively in networks with higher user densities
and with less power-efficient backhaul links.
\end{enumerate}

\subsection{Organization and Notations}
The rest of the paper is organized as follows. Section~\ref{Sec:SystemModel} presents the system model. Section~\ref{Sec:Formulation_Analysis} provides the problem formulation and problem analysis. In Section~\ref{Sec:LGSBF}, the LGSBF framework is proposed to minimize the network power consumption. Simulation results are demonstrated in Section~\ref{Sec:Simulation}. Finally, Section~\ref{Sec:Conclusions} concludes the paper.

Throughout this paper, vectors and matrices are denoted by lower-case and upper-case bold letters, respectively. The $\ell_{p}$-norm is represented by $\left\Vert \cdot\right\Vert _{p}$. The indicator function is denoted as $\boldsymbol{I}\left(\cdot\right)$, where $\boldsymbol{I}\left(e\right)=1$ if event $e$ is true, and
$\boldsymbol{I}\left(e\right)=0$ otherwise. We use $\left(\cdot\right)^{\mathsf{T}}$, $\left(\cdot\right)^{\mathsf{H}}$, $\mathrm{Tr}\left(\cdot\right)$ and $\mathrm{Re}\left\{ \cdot\right\} $ to denote transpose, Hermitian transpose, trace and real part operators, respectively. Calligraphy letters are used to denote sets.

\section{System Model} \label{Sec:SystemModel}
In this section, we will introduce the communication model, caching and backhaul models, as well as the power consumption model. Then the main performance metrics will be presented.

\subsection{Communication Model}
\begin{figure}
\centering \includegraphics[width=0.45\textwidth]{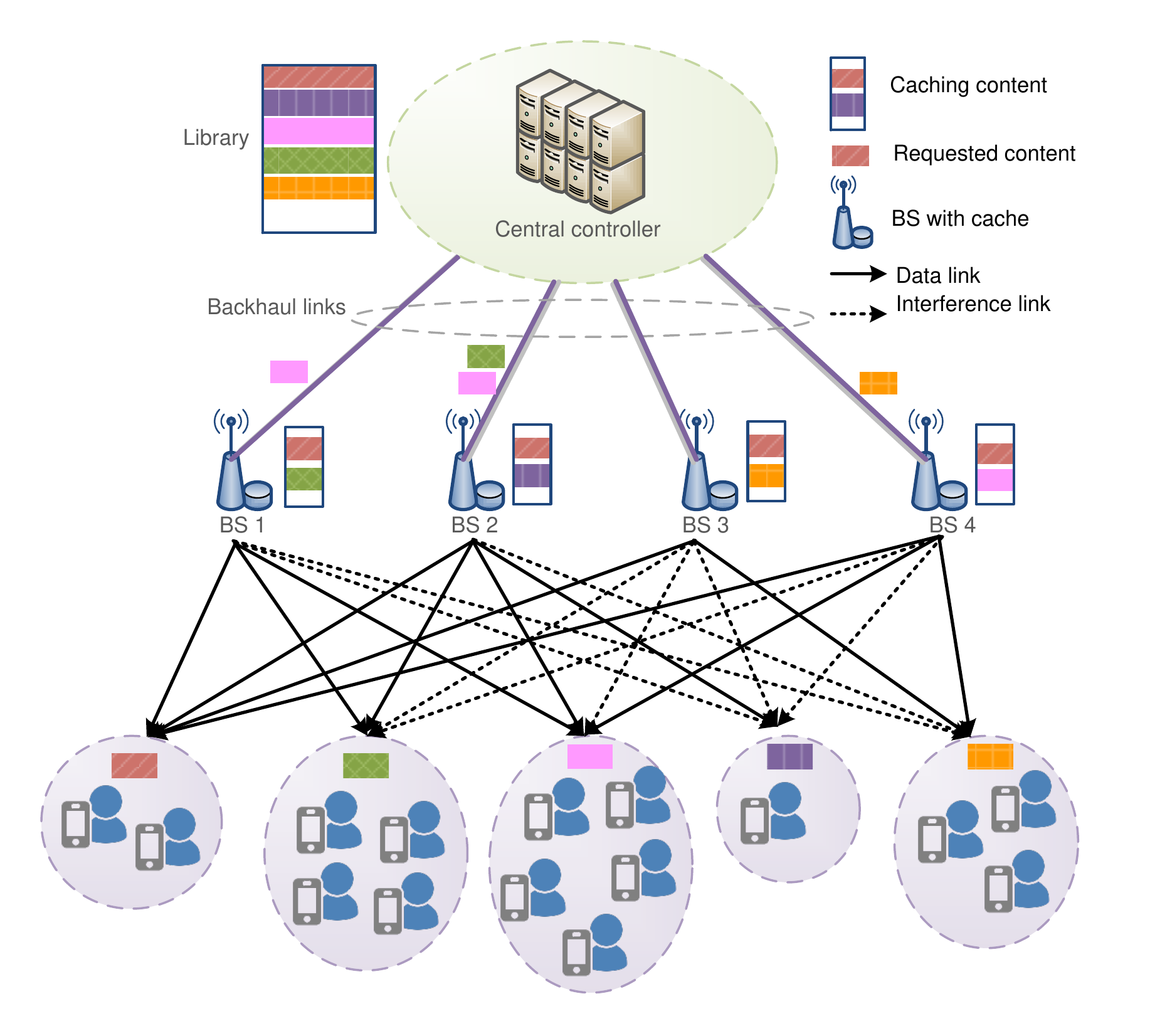}  
\caption{\footnotesize{}{}System model. MUs requesting the same content form a group served by a cluster of BSs via multicast transmission. The requested content is either cached at serving BSs or retrieved from the central controller via corresponding backhaul links.} \label{model}
\end{figure}

We consider a downlink multicast network consisting of $N_{U}$ single-antenna MUs cooperatively served by $N_{B}$ multi-antenna BSs, where the $j$-th BS has $L_{j}$ antennas. Each BS is equipped with a cache storage and connected to the central controller via a capacity-limited backhaul link. The central controller has access to the whole data library containing $N_{F}$ pieces of equal-size content objects. Let $\mathcal{J}=\left\{ 1,\dots,N_{B}\right\} $, $\mathcal{K}=\left\{ 1,\dots,N_{U}\right\} $ and $\mathcal{F}=\left\{ 1,\dots,N_{F}\right\} $ denote the sets of BSs, MUs and content objects, respectively. At the beginning of each interval, each MU makes a content request which follows a content popularity distribution. The MUs requesting the same content are grouped together and served by a cluster of BSs using multicast transmission. During each interval, the number of multicast groups is $N_{G}$ ($1\leq N_{G}\leq\min\left\{ N_{U},N_{F}\right\} $), and the set of groups is denoted as $\mathcal{M}=\left\{ 1,\dots,N_{G}\right\} $. The set of MUs in group $m$ is denoted as $\mathcal{G}_{m},\forall m\in\mathcal{M}$. Since each MU is assumed to request one piece of content during an interval, we have $\mathcal{G}_{m}\cap\mathcal{G}_{i}=\emptyset$, for $m\neq i$, and $\sum_{m=1}^{N_{G}}\left|\mathcal{G}_{m}\right|=N_{U}$. When BS $j$ caches the content requested by group $m$, BS $j$ can directly transmit the local content to group $m$. Otherwise, the uncached content has to be retrieved from the central controller to BS $j$ via the corresponding backhaul link and then transmitted to group $m$. The system model is illustrated in Fig. \ref{model}. 
The propagation channel from the $j$-th BS to the $k$-th MU is denoted as $\mathbf{h}_{kj}\in\mathbb{C}^{L_{j}},\forall k,j$, and the transmit beamforming vector from the $j$-th BS to the multicast group $m$ is denoted as $\mathbf{v}_{jm}\in\mathbb{C}^{L_{j}},\forall j,m$. The transmit signal at the $j$-th BS is given by\textcolor{black}{
	\begin{equation}
	\mathbf{x}_{j}=\sum_{m=1}^{N_{G}}\mathbf{v}_{jm}s_{m},\label{eq:tx signal}
	\end{equation}
}where $s_{m}\in\mathbb{C}$ stands for the encoded information symbol for the multicast group $m$ with $\mathbb{E}\left[\left|s_{m}\right|^{2}\right]=1$. The received signal at the MU $k\in\mathcal{G}_{m}$ is given by
\begin{equation}
y_{km}=\sum_{j=1}^{N_{B}}\mathbf{h}_{kj}^{\mathsf{H}}\mathbf{v}_{jm}s_{m}+\sum_{i=1,i\neq m}^{N_{G}}\sum_{j=1}^{N_{B}}\mathbf{h}_{kj}^{\mathsf{H}}\mathbf{v}_{ji}s_{i}+n_{k},\forall k\in\mathcal{G}_{m},\forall m\in\mathcal{M},\label{eq:Rx signal}
\end{equation}
where $n_{k}\sim\mathcal{CN}\left(0,\sigma_{k}^{2}\right)$ is the additive Gaussian noise at the $k$-th MU. Assume that all MUs adopt single user detection and thus treat interference as noise. The signal-to-interference-plus-noise
\textcolor{black}{ratio} (SINR) at MU $k\in\mathcal{G}_{m}$ is given by
\begin{equation}
\mathrm{SINR}_{k}=\frac{\left|\mathbf{h}_{k}^{\mathsf{H}}\mathbf{v}_{m}\right|^{2}}{\sum_{i\neq m}^{N_{G}}\left|\mathbf{h}_{k}^{\mathsf{H}}\mathbf{v}_{i}\right|^{2}+\sigma_{k}^{2}},\forall k\in\mathcal{G}_{m},\forall m\in\mathcal{M},\label{eq:SINR}
\end{equation}
where $\mathbf{h}_{k}=\left[\mathbf{h}_{k1}^{\mathsf{H}},\mathbf{h}_{k2}^{\mathsf{H}},\dots,\mathbf{h}_{kN_{B}}^{\mathsf{H}}\right]^{\mathsf{H}}\in\mathbb{C}^{N}$
with $N=\sum_{j=1}^{N_{B}}L_{j}$, represents the channel vector from all the BSs to the $k$-th MU, and $\mathbf{v}_{m}=\left[\mathbf{v}_{1m}^{\mathsf{H}},\mathbf{v}_{2m}^{\mathsf{H}},\dots,\mathbf{v}_{N_{B}m}^{\mathsf{H}}\right]^{\mathsf{H}}\in\mathbb{C}^{N}$ represents the beamforming vector from all the BSs to group $m$. Let $\mathbf{v}=\left[\tilde{\mathbf{v}}_{j}\right]_{j=1}^{N_{B}}\in\mathbb{C}^{N_{G}N}$
denote the aggregate beamforming vector with $\tilde{\mathbf{v}}_{j}=\left[\mathbf{v}_{jm}\right]_{m=1}^{N_{G}}\in\mathbb{C}^{N_{G}L_{j}}$
as the beamforming vector from the $j$-th BS to all multicast groups, i.e.,
\begin{equation}
\mathbf{v}=\left[\underset{\tilde{\mathbf{v}}_{1}^{H}}{\underbrace{\mathbf{v}_{11}^{H},\mathbf{v}_{12}^{H},\dots,\mathbf{v}_{1N_{G}}^{H}}},\dots,\underset{\tilde{\mathbf{v}}_{j}^{H}}{\underbrace{\mathbf{v}_{j1}^{H},\dots,\mathbf{v}_{jm}^{H},\dots,\mathbf{v}_{jN_{G}}^{H}}},\dots,\underset{\tilde{\mathbf{v}}_{N_{B}}^{H}}{\underbrace{\mathbf{v}_{N_{B}1}^{H},\mathbf{v}_{N_{B}2}^{H},\dots,\mathbf{v}_{N_{B}N_{G}}^{H}}}\right]^{H}.\label{eq:aggregate_v}
\end{equation}
To keep the analysis simple, we assume that each BS has the same number of antennas, i.e., $L_{j}=L,\forall j\in\mathcal{J}$. Define the target SINR vector as $\mathbf{\gamma}=\left[\gamma_{1},\dots,\gamma_{N_{G}}\right]$, where $\gamma_{m}$ stands for the lowest received SINR threshold for the users in group $m$. In order to decode the message successfully, any user $k\in{\cal G}_{m}$, should satisfy the following QoS constraint
\begin{equation}
\mathrm{SINR}_{k}\geq\gamma_{m},\forall k\in{\cal G}_{m},\forall m.\label{eq:QoS constraint}
\end{equation}
Denote the maximum transmit power of the $j$-th BS as $P_{j}^{\mathrm{TX}}$, and transmit power constraints are given by 
\begin{equation}
\sum_{m=1}^{N_{G}}\left\Vert \mathbf{v}_{jm}\right\Vert _{2}^{2}\leq P_{j}^{\mathrm{TX}},\forall j\in\mathcal{J}.\label{eq:tx power constaint}
\end{equation}

\subsection{Caching and Backhaul Models}
Caching networks operate in two phases, i.e., the prefetching phase and the delivery phase. In the prefetching phase, BSs fetch some contents from the file library of the central controller and store them at local caches, which usually happens during off-peak time. In the delivery phase (usually the busy \textcolor{black}{hours}), MUs may request arbitrary content in the file library. Since some desired contents have already been cached locally in the prefetching phase, only the rest of the requested content objects need to be delivered to BSs via backhaul links.

Define a caching matrix $\mathbf{C}=\left[c_{f,j}\right]\in\left\{ 0,1\right\} ^{N_{F}\times N_{B}}$, where $c_{f,j}=1$ means that the $f$-th content is cached at the $j$-th BS. Assume that MUs in group $m$ request content $q_{m}\in\left\{ 1,\dots,N_{F}\right\} $, and $c_{q_{m},j}=1$ means that the content requested by MUs in group $m$ is cached at the $j$-th BS. The transmit association status matrix is denoted as $\mathbf{T}=\left[t_{jm}\right]\in\left\{ 0,1\right\} ^{N_{B}\times N_{G}}$, where $t_{jm}=1$ means that the $j$-th BS serves group $m$ and $t_{jm}=0$ means the opposite. Let $\mathbf{N}^\mathrm{BA}=\left[n_{jm}\right]\in\left\{ 0,1\right\} ^{N_{B}\times N_{G}}$ denote the backhaul data assignment matrix, where $n_{jm}=1$ means that the content requested by the $m$-th user group will be assigned to the $j$-th BS via its backhaul link. It is not difficult to obtain that
\begin{equation}
n_{jm}=t_{jm}\left(1-c_{q_{m},j}\right).\label{eq:N-T-C}
\end{equation}
Therefore, only when $t_{jm}=1$ and $c_{q_{m},j}=0$, it will spawn backhaul traffic to retrieve the requested content, i.e.,
$n_{jm}=1$, and otherwise we have $n_{jm}=0$.

For ease of discussion, we consider fixed and feasible target SINR requirements as in \cite{Tao2015arXiv}. The transmission data rate for group $m$ is given by $R_{m}=B_{0}\log_{2}\left(1+\gamma_{m}\right)\,\left(\mathrm{bps}\right),$ where $B_{0}$ is the available bandwidth. The data rate (i.e., the traffic load) of backhaul link $j$ is then given by
\begin{equation}
R_{j}^{\mathrm{BH}}=\sum_{m=1}^{N_{G}}R_{m}n_{jm}\,\left(\mathrm{bps}\right),\forall j\in\mathcal{J}.\label{eq:bh_load_n}
\end{equation}
Since the capacity of each backhaul link is limited, we consider the following backhaul capacity constraints
\begin{equation}
R_{j}^{\mathrm{BH}}\leq C_{j}^{\mathrm{BH}},\forall j\in\mathcal{J}.\label{eq:BH-constraint}
\end{equation}

\subsection{Power Consumption Model}
We focus on the network power consumption of the delivery phase, for which the signal processing and optimization are much more challenging than the prefetching phase. Owing to the advances in caching hardwares \cite{CacheEnergyEfficiency}, caches have been made very energy-efficient. \textcolor{black}{Moreover, once the cache placement is finished in the prefetching phase, it will remain unchanged for a period of time, e.g., several days or weeks, since the file popularity evolves slowly, while user requests happen much more frequently. Therefore, the frequent reuse of cached contents can save considerable backhaul power consumption, which makes the power consumption in the prefetching phase negligible. Also, the cache placement is usually conducted during off-peak hours when the electricity resource is abundant and with a low price. Therefore, we focus on the network power consumption for the delivery phase, and omit the power consumption for caching.}  The main components of network power consumption, i.e, BS power consumption and backhaul power consumption, will be modeled as follows.

\subsubsection{BS Power Consumption Model}

We adopt the empirical linear model \cite{BS_PowerModel} to describe
the power consumption of the $j$-th BS: 
\begin{equation}
P_{j}^{\mathrm{BS}}=\begin{cases}
P_{A,j}^{\mathrm{BS}}+\delta_{j}P_{j}^{{\rm out}}, & \textrm{if }0<P_{j}^{{\rm out}}\leq P_{j}^{\mathrm{TX}}\\
P_{S,j}^{\mathrm{BS}}, & \textrm{if }P_{j}^{{\rm out}}=0
\end{cases},\label{eq:-8}
\end{equation}
where $P_{A,j}^{\mathrm{BS}}$ ($P_{S,j}^{\mathrm{BS}}$) stands for
the active (sleep) mode power consumption, $\delta_{j}$ represents
the slope of the load-dependent power consumption, and $P_{j}^{{\rm out}}$
is the BS transmit power, i.e., $P_{j}^{{\rm out}}=\sum_{m=1}^{N_{G}}\left\Vert \mathbf{v}_{jm}\right\Vert _{2}^{2}=\left\Vert \tilde{\mathbf{v}}_{j}\right\Vert _{2}^{2}$. Although a BS's power consumption can be arbitrarily close
to zero Watt in the deepest sleep level, it may cause an undesirable
long delay to wake up the BS from this low power mode \cite{sleepmode}.
In practice, when a BS has no transmission tasks, a less deep sleep
mode is usually adopted, where only some well-selected parts of the
hardware may be inactivated, in order to fasten the activation process.
As a result, it is typical to have $P_{S,j}^{\mathrm{BS}}\neq0$. For instance, according to the survey on BS power
consumption \cite{BS_PowerModel}, for a 2-antenna pico-BS, the typical
values are $P_{A,j}^{\mathrm{BS}}=6.8\textrm{ }\mathrm{W}$, $P_{S,j}^{\mathrm{BS}}=4.3\textrm{ }\mathrm{W}$
and $\delta_{j}=4$. Let $\mathcal{A}\subseteq{\cal J}$ and $\mathcal{Z}\subseteq{\cal J}$
denote the sets of active BSs and inactive BSs, respectively. Then,
the total BS power consumption is given by
\begin{equation}
\hat{p}_{1}=\sum_{j\in\mathcal{A}}\left(P_{A,j}^{\mathrm{BS}}+\delta_{j}\sum_{m=1}^{N_{G}}\left\Vert \mathbf{v}_{jm}\right\Vert _{2}^{2}\right)+\sum_{j\in\mathcal{Z}}P_{S,j}^{\mathrm{BS}}.\label{eq:bs-power}
\end{equation}
Based on the BS power consumption model, we conclude that it is essential
to put BSs into sleep mode whenever possible in order to save the
power consumption.

\subsubsection{Backhaul Transport Power Consumption Model}

The total backhaul transport power consumption is given by
\begin{equation}
\hat{p}_{2}=\sum_{j=1}^{N_{B}}P_{j}^{\mathrm{BH}},\label{eq:-2}
\end{equation}
where $P_{j}^{\mathrm{BH}}$ is the power consumption of the backhaul
link corresponding to BS $j$. Similar to the BS power consumption model, we need to consider both
active and sleep modes for backhaul links. The power consumption of an active backhaul link turns out to be   traffic-dependent \cite{EngergyEfficiencyBH}. Therefore, the backhaul transport power consumption is expressed as 
\begin{align}
P_{j}^{\mathrm{BH}} & =\begin{cases}
P_{A,j}^{\mathrm{BH}}+\frac{R_{j}^{\mathrm{BH}}}{C_{j}^{\mathrm{BH}}}P_{j}^{\mathrm{max}}, & \textrm{if }0<R_{j}^{\mathrm{BH}}\leq C_{j}^{\mathrm{BH}}\\
P_{S,j}^{\mathrm{BH}}, & \textrm{if }R_{j}^{\mathrm{BH}}=0
\end{cases},\forall j\in\mathcal{J},\label{eq:-3}
\end{align}
where $C_{j}^{\mathrm{BH}}$ denotes the maximum data rate (i.e.,
capacity) of the backhaul link, $P_{j}^{\mathrm{max}}$ represents the
backhaul power consumption when supporting the maximum data rate,
and $E_{j}^{\mathrm{BH}}\triangleq P_{j}^{\mathrm{max}}\left/C_{j}^{\mathrm{BH}}\right.$
is the backhaul transport energy coefficient. For a backhaul link, typical values are
$P_{A,j}^{\mathrm{BH}}=3.85\textrm{ }\mathrm{W}$, $P_{S,j}^{\mathrm{BH}}=0.75\textrm{ }\mathrm{W}$.
The typical value for $E_{j}^{\mathrm{BH}}$ is around $10^{-7}\,\mathrm{J/bit}$
for microwave backhaul link \cite{EngergyEfficiencyBH}, and around
$10^{-5}\,\mathrm{J/bit}$ for copper DSL \cite{DSL_value}.
The power consumption of all backhaul links can be calculated as
\begin{equation}
\hat{p}_{2}=\sum_{j\in\mathcal{A}}\left(P_{A,j}^{\mathrm{BH}}+E_{j}^{\mathrm{BH}}R_{j}^{\mathrm{BH}}\right)+\sum_{j\in\mathcal{Z}}P_{S,j}^{\mathrm{BH}}.\label{eq:bh-power}
\end{equation}
 Combining formula (\ref{eq:bh_load_n}), (\ref{eq:bs-power})
and (\ref{eq:bh-power}), we have the total network power consumption
as
\begin{align}
\tilde{p}\left(\mathcal{A},\mathbf{T},\mathbf{v}\right) & =\hat{p}_{1}+\hat{p}_{2}\label{eq:-4}\\
 & =\sum_{j\in\mathcal{A}}\delta_{j}\sum_{m=1}^{N_{G}}\left\Vert \mathbf{v}_{jm}\right\Vert _{2}^{2}+\sum_{j\in\mathcal{A}}\sum_{m=1}^{N_{G}}E_{j}^{\mathrm{BH}}R_{m}n_{jm}+\sum_{j\in\mathcal{A}}P_{j}^{\mathrm{D}}+\sum_{j\in\mathcal{J}}\left(P_{S,j}^{\mathrm{BS}}+P_{S,j}^{\mathrm{BH}}\right),\label{eq:power-overall}
\end{align}
where
\begin{equation}
P_{j}^{\mathrm{D}}=\left(P_{A,j}^{\mathrm{BS}}-P_{S,j}^{\mathrm{BS}}\right)+\left(P_{A,j}^{\mathrm{BH}}-P_{S,j}^{\mathrm{BH}}\right)\label{eq:P_D}
\end{equation}
is the difference of static state power consumption between active
and sleep modes for BS $j$ and its corresponding backhaul link, and
is named as the \emph{relative power consumption} for simplification.
As a matter of fact, usually we have $P_{A,j}^{\mathrm{BS}}>P_{S,j}^{\mathrm{BS}}$
and $P_{A,j}^{\mathrm{BH}}>P_{S,j}^{\mathrm{BH}}$, and thus $P_{j}^{\mathrm{D}}>0$. Let $\beta_{jm}=E_{j}^{\mathrm{BH}}R_{m}$ denote the backhaul power consumption for BS $j$ for serving user group $m$.
Since constant terms will not influence the optimization design, we
can equivalently minimize the re-defined network power consumption
instead of (\ref{eq:power-overall}):
\begin{equation}
p\left(\mathcal{A},\mathbf{N}^\mathrm{BA},\mathbf{v}\right)=\sum_{j\in\mathcal{A}}\delta_{j}\sum_{m=1}^{N_{G}}\left\Vert \mathbf{v}_{jm}\right\Vert _{2}^{2}+\sum_{j\in\mathcal{A}}\sum_{m=1}^{N_{G}}\beta_{jm}n_{jm}+\sum_{j\in\mathcal{A}}P_{j}^{\mathrm{D}},\label{eq:Power_reDefined}
\end{equation}
which consists of BS transmit power consumption, traffic-dependent backhaul power consumption, and relative power consumption of active BSs and corresponding backhaul links.

\section{Problem Formulation and Analysis}

\label{Sec:Formulation_Analysis}

In this section, we will first formulate the network power minimization problem, which will then be analyzed and reformulated to reveal the layered group sparsity structure in the optimization variables. Based on (\ref{eq:Power_reDefined}), there are three strategies minimizing the network power consumption: i) to reduce the relative power consumption by switching off as many BSs and corresponding backhaul links as possible; ii) to reduce the transmit power consumption of BSs with coordinated beamforming by having more active BSs; and iii) to reduce the traffic-dependent backhaul power consumption by minimizing backhaul delivery of uncached content. Obviously, these strategies cannot be achieved at the same time. Hence, the network power consumption minimization problem will be a joint design across BS selection, backhaul data assignment and coordinated transmit beamforming.

\subsection{Problem Formulation}
In this work, we assume that perfect channel state information (CSI) $\left\{ \mathbf{h}_{k}\right\} $, cache placement $\mathbf{C}$, and  overall user requests $\left\{ c_{q_{m},j}\right\} $ are known a priori at the central controller.
Considering MU QoS requirements, BS transmit power constraints and per-BS backhaul capacity constraints, we formulate the network power consumption minimization problem as a joint active BS selection, backhaul data assignment and transmit beamforming design problem:
\begin{align}
\mathscr{P}\textrm{: }\underset{\mathcal{A},\left\{ n_{jm}\right\} ,\left\{ \mathbf{v}_{jm}\right\} }{\mathrm{minimize}}\mathrm{\quad} & p\left(\mathcal{A},\mathbf{N}^\mathrm{BA},\mathbf{v}\right)\label{eq:prob_p_original}\\
\textrm{subject to\quad} & \frac{\left|\mathbf{h}_{k}^{\mathsf{H}}\mathbf{v}_{m}\right|^{2}}{\sum_{i\neq m}^{N_{G}}\left|\mathbf{h}_{k}^{\mathsf{H}}\mathbf{v}_{i}\right|^{2}+\sigma_{k}^{2}}\geq\gamma_{m},\forall k\in\mathcal{G}_{m},\forall m\in\mathcal{M}\tag{\ref{eq:prob_p_original}a}\nonumber \\
 & \sum_{m=1}^{N_{G}}\left\Vert \mathbf{v}_{jm}\right\Vert _{2}^{2}\leq P_{j}^{\mathrm{TX}},\forall j\in\mathcal{J}\tag{\ref{eq:prob_p_original}b}\nonumber \\
 & \sum_{m=1}^{N_{G}}R_{m}n_{jm}\leq C_{j}^{\mathrm{BH}},\forall j\in\mathcal{J}\tag{\ref{eq:prob_p_original}c}\nonumber \\
 & \mathbf{N}^\mathrm{BA}=\left[n_{jm}\right]\in\left\{ 0,1\right\} ^{N_{B}\times N_{G}}.\tag{\ref{eq:prob_p_original}d}\nonumber 
\end{align}
In the following subsection, we will analyze problem $\mathscr{P}$, which will motivate us to reformulate it for developing low-complexity algorithms.

\subsection{Problem Analysis}
In this subsection, we will identify the main challenges of the network
power minimization problem $\mathscr{P}$. We first consider the case
with a given active BS set $\mathcal{A}$ and a given backhaul data
assignment matrix $\mathbf{N}^\mathrm{BA}$, resulting in a transmit power
minimization problem given by
\begin{align}
\mathscr{P}\left(\mathcal{A},\mathbf{N}^\mathrm{BA}\right)\textrm{: }\underset{\left\{ \mathbf{v}_{jm}\right\} }{\mathrm{minimize}}\mathrm{\quad} & \sum_{j\in\mathcal{A}}\delta_{j}\sum_{m=1}^{N_{G}}\left\Vert \mathbf{v}_{jm}\right\Vert _{2}^{2}\label{eq:eq:prob-p-A-N}\\
\textrm{subject to\quad} & \left(\ref{eq:prob_p_original}\mathrm{a}\right),\left(\ref{eq:prob_p_original}\mathrm{b}\right),\nonumber 
\end{align}
which is a multicast beamforming problem as discussed in \cite{multicasting}. 

The above analysis implies that once the optimal $\mathcal{A}$ and
$\mathbf{N}^\mathrm{BA}$ are identified, the solution $\mathbf{v}$ can
be determined by solving problem $\mathscr{P}\left(\mathcal{A},\mathbf{N}^\mathrm{BA}\right)$.
Thus, problem $\mathscr{P}$ can be
solved by searching over all the possible active BS sets and all possible
$\mathbf{N}^\mathrm{BA}$'s, i.e., 
\begin{equation}
p^{\star}=\underset{\begin{array}{c}
Q\in\left\{ A,\dots,N_{B}\right\} \end{array}}{\mathrm{minimize}}p^{\star}\left(Q\right),\label{eq:-5}
\end{equation}
where $A\geq1$ is the minimum number of active BSs to meet the QoS
constraints, and $p^{\star}\left(Q\right)$ is determined by
\begin{equation}
p^{\star}\left(Q\right)=\underset{\begin{array}{c}
\mathcal{A}\subseteq{\cal J},\left|\mathcal{A}\right|=Q\\
\mathbf{N}^\mathrm{BA}\in\left\{ 0,1\right\} ^{N_{B}\times N_{G}}
\end{array}}{\mathrm{minimize}}p^{\star}\left({\cal A},\mathbf{N}^\mathrm{BA}\right),\label{eq:-6}
\end{equation}
where $p^{\star}\left({\cal A},\mathbf{N}^\mathrm{BA}\right)$ is the optimal
value of problem $\mathscr{P}\left(\mathcal{A},\mathbf{N}^\mathrm{BA}\right)$
and $\left|\mathcal{A}\right|$ is the cardinality of set ${\cal A}$.
Since the number of subsets ${\cal A}$ of size $a$ is
$\binom{N_{B}}{a}$ and we need to search over $2^{N_{B}N_{G}}$ possible
$\mathbf{N}^\mathrm{BA}$'s for each subset ${\cal A}$, the complexity of
the overall search procedure will grow exponentially with $N_{B}\left(N_{G}+1\right)$,
which makes this approach unscalable. Therefore, the key to solve the problem
is to effectively determine ${\cal A}^{\star}$ and $\mathbf{N}^\mathrm{BA\star}$.
This problem needs to be reformulated to develop more efficient
algorithms.

\subsection{Layered Group Sparse Beamforming Formulation}

In the following, we will reformulate the original problem. First, let us present several key observations, aiming at exploiting the unique structure of the problem, which will help us address the main challenges. The original objective can be decomposed into three parts, i.e.,
\begin{equation}
p\left(\mathcal{A},\mathbf{N}^\mathrm{BA},\mathbf{v}\right)=T\left(\mathbf{v}\right)+F_{1}\left(\mathcal{A}\right)+F_{2}\left(\mathcal{A},\mathbf{N}^\mathrm{BA}\right),\label{eq:-15}
\end{equation}
where $T\left(\mathbf{v}\right)=\sum_{j=1}^{N_{B}}\sum_{m=1}^{N_{G}}\delta_{j}\left\Vert \mathbf{v}_{jm}\right\Vert _{2}^{2}$ is the BS transmit power consumption, $F_{1}\left(\mathcal{A}\right)=\sum_{j\in\mathcal{A}}P_{j}^{\mathrm{D}}$
is the relative power consumption, and $F_{2}\left(\mathcal{A},\mathbf{N}^\mathrm{BA}\right)=\sum_{j\in\mathcal{A}}\sum_{m=1}^{N_{G}}\beta_{jm}n_{jm}$ is the backhaul power consumption. We will show that $F_{1}$ and $F_{2}$ can be expressed as functions of the aggregate beamforming
vector $\mathbf{v}$, which are able to indicate the group sparsity of $\mathbf{v}$ at different layers. 

\subsubsection{BS-layer Group Sparsity of $\mathbf{v}$} All the coefficients in a given vector $\tilde{\mathbf{v}}_{j}=\left[\mathbf{v}_{jm}\right]_{m=1}^{N_{G}}\in\mathbb{C}^{N_{G}L_{j}}$
form a \emph{BS-layer} group and $\sum_{j=1}^{N_{B}}\boldsymbol{I}\left(\left\Vert \mathbf{\tilde{v}}_{j}\right\Vert _{2}>0\right)$ can be considered as a group sparsity measure of $\mathbf{v}$. When
the $j$-th BS is switched off, all the coefficients in vector $\tilde{\mathbf{v}}_{j}$
will be set to zero, i.e., $\tilde{\mathbf{v}}_{j}=\mathbf{0}$. It
is possible that multiple BSs can be switched off and the corresponding
beamformers will be set to zero, which means that $\mathbf{v}$ has
a \emph{BS-layer} group sparsity structure. It is observed that if
$\left\Vert \mathbf{\tilde{v}}_{j}\right\Vert _{2}>0$, then we have
$j\in\mathcal{A}$, and if $\left\Vert \mathbf{\tilde{v}}_{j}\right\Vert _{2}=0$,
we have $j\in\mathcal{Z}$. Therefore, for a given beamformer $\mathbf{v}$, the relative power consumption $F_{1}\left(\mathcal{A}\right)$ can be rewritten as 
\begin{equation}
F_{1}\left(\mathbf{v}\right)=\sum_{j=1}^{N_{B}}P_{j}^{\mathrm{D}}I\left(\left\Vert \mathbf{\tilde{v}}_{j}\right\Vert _{2}>0\right).\label{eq:F1-supp}
\end{equation}

\subsubsection{Data Assignment-layer Group Sparsity of $\mathbf{v}$} The backhaul data assignment matrix $\mathbf{N}^\mathrm{BA}$ can be fully specified with the knowledge of the beamformer $\mathbf{v}$ as 
\begin{equation}
n_{jm}=\left(1-c_{q_{m},j}\right)\boldsymbol{I}\left(\left\Vert \mathbf{v}_{jm}\right\Vert _{2}>0\right).\label{eq:n-indicator-func}
\end{equation}
From (\ref{eq:n-indicator-func}), we observe that for a given
user group $m$, when BS $j$ does not serve it, i.e., $\mathbf{v}_{jm}=\mathbf{0}$,
there is no need to assign the content requested by this user group
to BS $j$, and $n_{jm}$ will be set to zero; when the content requested
by user group $m$ happens to be cached at BS $j$, i.e., $c_{q_{m},j}=1$,
regardless of whether BS $j$ serves this user group or not, there
is no need to assign the content requested by this user group to BS
$j$, and hence $n_{jm}$ will always be set to zero. It is likely
that we can reduce the number of backhaul data assignments and the
corresponding $n_{jm}$ values will be set to zero, from which we can infer
that the backhaul data assignment matrix $\mathbf{N}^\mathrm{BA}$ has a
sparsity structure. In addition, we observe 
\begin{equation}
\left\Vert \mathbf{N}^{BA}\right\Vert _{0}\leq\sum_{j=1}^{N_{B}}\sum_{m=1}^{N_{G}}\boldsymbol{I}\left(\left\Vert \mathbf{v}_{jm}\right\Vert _{2}>0\right),\label{eq:-7}
\end{equation}
which means that minimizing $\sum_{j=1}^{N_{B}}\sum_{m=1}^{N_{G}}\boldsymbol{I}\left(\left\Vert \mathbf{v}_{jm}\right\Vert _{2}>0\right)$  can imply the minimization of $\left\Vert \mathbf{N}^\mathrm{BA}\right\Vert _{0}$. All the coefficients in a given vector $\mathbf{v}_{jm}\in\mathbb{C}^{L_{j}}$
form a group and $\sum_{j=1}^{N_{B}}\sum_{m=1}^{N_{G}}\boldsymbol{I}\left(\left\Vert \mathbf{v}_{jm}\right\Vert _{2}>0\right)$
can be considered as another group sparsity measure of $\mathbf{v}$.
Since this measure is related to the backhaul data assignment, it
can be regarded to represent the ``\textit{data assignment-layer}''
group sparsity. Hence, the backhaul power consumption $F_{2}\left(\mathcal{A},\mathbf{N}^\mathrm{BA}\right)$
can be rewritten as
\begin{equation}
F_{2}\left(\mathbf{v}\right)=\sum_{j=1}^{N_{B}}\sum_{m=1}^{N_{G}}\beta_{jm}\left(1-c_{q_{m},j}\right)\boldsymbol{I}\left(\left\Vert \mathbf{v}_{jm}\right\Vert _{2}>0\right)I\left(\left\Vert \mathbf{\tilde{v}}_{j}\right\Vert _{2}>0\right).\label{eq:F2-supp-1}
\end{equation}
For a given BS $j$, $\left\{ \mathbf{v}_{jm}\right\} $ are non-overlapping
subgroups of $\mathbf{\tilde{v}}_{j}$, and $\left\Vert \mathbf{v}_{jm}\right\Vert _{2}>0$
is a sufficient condition for $\left\Vert \mathbf{\tilde{v}}_{j}\right\Vert _{2}>0$.
As a result, $F_{2}\left(\mathbf{v}\right)$ can be simplified as
\begin{equation}
F_{2}\left(\mathbf{v}\right)=\sum_{j=1}^{N_{B}}\sum_{m=1}^{N_{G}}\beta_{jm}\left(1-c_{q_{m},j}\right)\boldsymbol{I}\left(\left\Vert \mathbf{v}_{jm}\right\Vert _{2}>0\right).\label{eq:F2-supp-2}
\end{equation}
Based on the above discussions, the network power minimization problem
$\mathscr{P}$ can be equivalently reformulated as the following group
sparse beamforming problem: 
\begin{align}
\mathscr{P}^{LGSBF}\textrm{: }\underset{\mathbf{v}}{\mathrm{minimize}}\mathrm{\quad} & p^{LGSBF}\left(\mathbf{v}\right)=T\left(\mathbf{v}\right)+F_{1}\left(\mathbf{v}\right)+F_{2}\left(\mathbf{v}\right)\label{eq:prob-p-LGSBF}\\
\textrm{subject to\quad} & \sum_{m=1}^{N_{G}}R_{m}\left(1-c_{q_{m},j}\right)\boldsymbol{I}\left(\left\Vert \mathbf{v}_{jm}\right\Vert _{2}>0\right)\leq C_{j}^{\mathrm{BH}},\forall j\in\mathcal{J},\tag{\ref{eq:prob-p-LGSBF}a}\nonumber \\
 & \left(\ref{eq:prob_p_original}\mathrm{a}\right),\left(\ref{eq:prob_p_original}\mathrm{b}\right).\nonumber 
\end{align}
The equivalence between problem $\mathscr{P}$ and problem $\mathscr{P}^{LGSBF}$
means that if $\mathbf{v}^{\star}$ is a solution to problem $\mathscr{P}^{LGSBF}$, then
$\left(\mathcal{A}^{\star},\left\{ n_{jm}^{\star}\right\} ,\left\{ \mathbf{v}_{jm}^{\star}\right\} \right)$
with $\mathcal{A}^{\star}=\left\{ j\left|\left\Vert \mathbf{\tilde{v}}_{j}^{\star}\right\Vert _{2}>0,j\in{\cal J}\right.\right\} $
and $n_{jm}^{\star}=\left(1-c_{q_{m},j}\right)\boldsymbol{I}\left(\left\Vert \mathbf{v}_{jm}^{\star}\right\Vert _{2}>0\right)$ is a solution to problem $\mathscr{P}$, and vice versa.

The incorporation of two group-sparsity measures in our problem formulation
generalizes those in previous works \cite{GSBF,Tao2015arXiv} which
considered only one group-sparsity measure. Notice that all these
group sparse beamforming problems can be unified in the following
generalized group structured optimization problem:
\begin{align}
\underset{\mathbf{v}}{\mathrm{minimize}}\mathrm{\quad} & T\left(\mathbf{v}\right)+\lambda_{1}\sum_{j=1}^{N_{B}}\alpha_{j}I\left(\left\Vert \mathbf{\tilde{v}}_{j}\right\Vert _{2}>0\right)+\lambda_{2}\sum_{j=1}^{N_{B}}\sum_{m=1}^{N_{G}}\eta_{jm}\boldsymbol{I}\left(\left\Vert \mathbf{v}_{jm}\right\Vert _{2}>0\right)\label{eq:general-lasso}\\
\textrm{subject to\quad} & \left(\ref{eq:prob_p_original}\mathrm{a}\right),\left(\ref{eq:prob_p_original}\mathrm{b}\right),\left(\ref{eq:prob-p-LGSBF}\mathrm{a}\right),\nonumber 
\end{align}
where $T\left(\mathbf{v}\right)=\sum_{j=1}^{N_{B}}\sum_{m=1}^{N_{G}}\delta_{j}\left\Vert \mathbf{v}_{jm}\right\Vert _{2}^{2}$
is a smoothed convex function, and $\lambda_{k}\geq0,\forall k\in\left\{ 1,2\right\} $
are regularization parameters for groups at different layers. When
$\left\{ \lambda_{k}\right\} $ \textcolor{black}{take variant combinations}, the model
falls into different problems, as shown in Table \ref{tab:general-lasso}.
\begin{table}
\centering\caption{Generalized Group Sparse Beamforming Model} \label{tab:general-lasso}
\begin{tabular}{c|c|c}
\hline 
Parameters & Problem & Algorithm\tabularnewline
\hline 
$\lambda_{1}=0,\lambda_{2}=0$ & Transmit power minimization & Coordinated beamforming \cite{MIMO_clusterLinearPrecoding}\tabularnewline
\hline 
$\lambda_{1}>0,\lambda_{2}=0$ & BS selection & \multirow{2}{*}{Group sparse beamforming \cite{HongLuoZQ2013,GSBF,Tao2015arXiv} }\tabularnewline
\cline{1-2} 
\multirow{1}{*}{$\lambda_{1}=0,\lambda_{2}>0$} & Backhaul data assignment & \tabularnewline
\hline 
$\lambda_{1}>0,\lambda_{2}>0$ & BS selection + backhaul data assignment & Proposed layered group sparse beamforming\tabularnewline
\hline 
\end{tabular}
\end{table}
In our formulation, we have $\lambda_{1}>0,\lambda_{2}>0$, which
means that we incorporate multiple sparsity-inducing regularizers
into the objective function, and therefore enable joint optimization
of BS selection and backhaul data assignment, which generalizes the
previous works. Specifically, the entries of $\mathbf{v}$ are partitioned into different
groups at two layers: i) the BS-layer where the beamforming coefficients
sent from each BS form a group (the number of groups of this layer
is $N_{B}$), and ii) the data assignment-layer where the beamforming
coefficients associated with one BS and one user group are considered
as a group (the number of groups of this layer is $N_{B}N_{G}$).
Furthermore, the previous works \cite{GSBF,Tao2015arXiv} failed to
take the per-BS backhaul capacity constraints into consideration,
which restricts their practical applications. In order to solve problem
$\mathscr{P}^{LGSBF}$, we are confronted with several unique challenges
which are highlighted as follows.

\subsubsection{Combinatorial Objective Function}

There are two indicator functions in $p^{LGSBF}\left(\mathbf{v}\right)$,
acting as two group-sparsity measures inducing group sparsity at different
layers to the problem. Moreover, the variables in the two group-sparsity
measures are non-separable. All the existing group sparse beamforming
methods \cite{GSBF,SparseBF_userCentricCluster,smoothedLp,Tao2015arXiv}
can only deal with one group-sparsity measure and are not applicable to our problem.

\subsubsection{Non-convex Quadratic QoS Constraints}

The non-convex quadratic QoS constraints are yielded by the physical-layer
multicast beamforming problem. Consider problem $\mathscr{P}\left(\mathcal{A},\mathbf{N}^\mathrm{BA}\right)$
as an example. On one hand, a ready approach to deal with these constraints
is to apply a semidefinite relaxation (SDR) technique \cite{multicasting},
and relax problem $\mathscr{P}\left(\mathcal{A},\mathbf{N}^\mathrm{BA}\right)$
into a semidefinite programming (SDP) problem by removing the rank-one
constraints, at the price of lifting the variables to higher dimensions.
On the other hand, the non-convex quadratic QoS constraints can also
be rewritten into the DC form \cite{Tao2015arXiv}. Compared to the
SDP transformation, the DC transformation will not incur loss of optimality
since it does not involve rank-one constraints. Moreover, the number
of variables in the SDP transformation is almost the square of that
in problem $\mathscr{P}\left(\mathcal{A},\mathbf{N}^\mathrm{BA}\right)$,
while the number of variables in the DC transformation nearly remains
the same as that in problem $\mathscr{P}\left(\mathcal{A},\mathbf{N}^\mathrm{BA}\right)$.

\subsubsection{Non-convex per-BS Backhaul Capacity Constraints}

Besides the aforementioned difficulties, the discrete indicator function
$\boldsymbol{I}\left(\left\Vert \mathbf{v}_{jm}\right\Vert _{2}>0\right)$
in per-BS backhaul capacity constraint, which characterizes whether
BS $j$ serves user group $m$, makes the problem much more challenging.
A key observation is that the indicator function can be equivalently
expressed as an $\ell_{0}$-norm of a scalar. The $\ell_{0}$-norm
stands for the number of nonzero entries in a vector, and reduces
to an indicator function in the scalar case. By using ideas from previous
literature \cite{SparseBF_userCentricCluster}, we may further approximate
the non-convex $\ell_{0}$-norm by a convex reweighted $\ell_{1}$-norm. 

Based on the challenges identified above, we will propose a low-complexity
algorithm to solve the problem efficiently based on the formulation
$\mathscr{P}^{LGSBF}$ in the following section.

\section{Layered Group Sparse Beamforming Framework} \label{Sec:LGSBF}
\begin{figure}
\centering{}\includegraphics[width=0.85\textwidth]{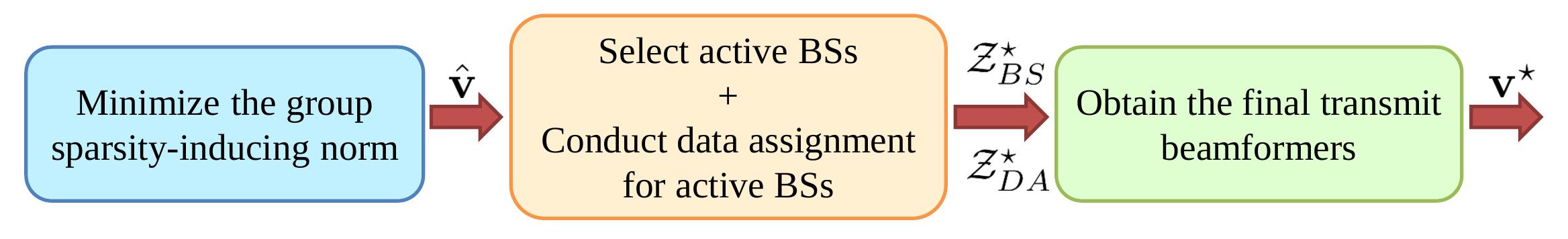}\caption{\label{fig:framework}Proposed generalized three-stage LGSBF framework.}
\end{figure}

In this section, based on the formulation $\mathscr{P}^{LGSBF}$,
we will develop a low-complexity algorithm. The main motivation is
to induce group sparsity in the aggregate beamformer $\mathbf{v}$
at both the BS-layer and the data assignment-layer to minimize the
total network power consumption. The proposed framework has three
stages, as shown in Fig. \ref{fig:framework}.
At the first stage, we solve a reweighted group sparsity-inducing
norm minimization problem, so as to induce a group sparsity structure in the
aggregate beamformer. Then, in the second stage, based on the approximately
sparse beamformer obtained from the first stage, we will conduct a
two-layer iterative search procedure, which can efficiently identify the
active BSs and backhaul data assignment, respectively. In the last
stage, with the knowledge of the active BS set and backhaul data assignment,
coordinated multicast beamforming will be adopted to obtain the final beamformers.
The details will be presented in the following subsections.

\subsection{Preliminaries and Motivation of LGSBF framework}
In order to induce group sparsity in the aggregate beamformer $\mathbf{v}$,
we first replace the indicator functions by $\ell_{0}$-norm, which
is thereafter relaxed into the mixed $\left.\ell_{1}\right/\ell_{p}$-norm
($p>1$) \cite{Bach2012}. The mixed $\left.\ell_{1}\right/\ell_{2}$-norm
and $\left.\ell_{1}\right/\ell_{\infty}$-norm are two commonly used
norms (also called regularizers) for inducing group sparsity. The
mixed $\left.\ell_{1}\right/\ell_{2}$-norm is the most common choice
and known as the group least-absolute selection and shrinkage operator
(group Lasso). In this study, we also adopt $p=2$, and obtain a convex
approximation for the objective $p^{LGSBF}\left(\mathbf{v}\right)$ as
\begin{align}
\widehat{p}\left(\mathbf{v}\right) & =\sum_{j=1}^{N_{B}}\sum_{m=1}^{N_{G}}\delta_{j}\left\Vert \mathbf{v}_{jm}\right\Vert _{2}^{2}+\sum_{j=1}^{N_{B}}P_{j}^{\mathrm{D}}\tilde{\omega}_{j}\left\Vert \mathbf{\tilde{v}}_{j}\right\Vert _{2}+\sum_{j=1}^{N_{B}}\sum_{m=1}^{N_{G}}\beta_{jm}\left(1-c_{q_{m},j}\right)\omega_{jm}\left\Vert \mathbf{v}_{jm}\right\Vert _{2},\label{eq:p_hat}
\end{align}
where $\left\{ \tilde{\omega}_{j}\right\} ,j=1,\dots,N_{B}$, and
$\left\{ \omega_{jm}\right\} ,j=1,\dots,N_{B},m=1,\dots,N_{G}$, are
positive weights. Compared with existing sparse beamforming methods
dealing with only one sparsity-inducing regularizer \cite{smoothedLp,Tao2015arXiv},
the problem $\mathscr{P}^{LGSBF}$ is even more
complicated since its objective function has incorporated two sparsity-inducing regularizers.
The multiple sparsity-inducing regularizers indicate that the solution
has a layered group sparse pattern, based on which we name the proposed
framework as a \emph{layered group sparse beamforming} (LGSBF) framework.
Moreover, in our problem, we are facing noncovex constraints as discussed in Section~\ref{Sec:Formulation_Analysis}, which
add more challenges.

\subsection{Stage I: Group Structured Sparsity Inducing Norm Minimization}

In this subsection, we propose a \textcolor{black}{convex relaxation} for problem $\mathscr{P}^{LGSBF}$.
To start with, we adopt the DC transformation to deal with the non-convex
quadratic QoS constraints, which are rewritten as
\begin{equation}
\gamma_{k}\left(\sum_{i\neq m}^{N_{G}}\left|\mathbf{h}_{k}^{\mathsf{H}}\mathbf{v}_{i}\right|^{2}+\sigma_{k}^{2}\right)-\left|\mathbf{h}_{k}^{\mathsf{H}}\mathbf{v}_{m}\right|^{2}\leq0,\forall k\in\mathcal{G}_{m},\forall m\in\mathcal{M}.\label{eq:QoS-DC}
\end{equation}
Then, we need to address the indicator functions in both the objective
function and backhaul capacity constraints. As stated in (\ref{eq:p_hat}),
we use the convex surrogate $\widehat{p}\left(\mathbf{v}\right)$
for the objective function $p^{LGSBF}\left(\mathbf{v}\right)$ by
employing the mixed $\left.\ell_{1}\right/\ell_{2}$-norm to approximate
the nonconvex $\ell_{0}$-norm. The indicator function
can be equivalently expressed as an $\ell_{0}$-norm of another scalar
$\left\Vert \mathbf{v}_{jm}\right\Vert _{2}^{2}$ instead of $\left\Vert \mathbf{v}_{jm}\right\Vert _{2}$, i.e.,
\begin{equation}
I\left(\left\Vert \mathbf{\tilde{v}}_{j}\right\Vert _{2}>0\right)=\left\Vert \left\Vert \mathbf{\tilde{v}}_{j}\right\Vert _{2}^{2}\right\Vert _{0},\textrm{ and }\boldsymbol{I}\left(\left\Vert \mathbf{v}_{jm}\right\Vert _{2}>\mathbf{0}\right)=\left\Vert \left\Vert \mathbf{v}_{jm}\right\Vert _{2}^{2}\right\Vert _{0},\label{eq:}
\end{equation}
which allows us to extend the mixed $\left.\ell_{1}\right/\ell_{2}$-norm
approximation. In order to further enhance sparsity, we employ an
iterative re-weighted $\left.\ell_{1}\right/\ell_{2}$-minimization,
inspired by the reweighted $\ell_{1}$-minimization proposed in \cite{Reweighted-L1_candes2008}.
The surrogate objective is rewritten as
\begin{equation}
\tilde{p}\left(\mathbf{v}\left|\tilde{\omega}_{j},\omega_{jm}\right.\right)=\sum_{j=1}^{N_{B}}\sum_{m=1}^{N_{G}}\delta_{j}\left\Vert \mathbf{v}_{jm}\right\Vert _{2}^{2}+\sum_{j=1}^{N_{B}}P_{j}^{\mathrm{D}}\tilde{\omega}_{j}\left\Vert \mathbf{\tilde{v}}_{j}\right\Vert _{2}^{2}+\sum_{j=1}^{N_{B}}\sum_{m=1}^{N_{G}}\beta_{jm}\left(1-c_{q_{m},j}\right)\omega_{jm}\left\Vert \mathbf{v}_{jm}\right\Vert _{2}^{2},\label{eq:-16}
\end{equation}
and the problem is reformulated as
\begin{align}
\mathscr{P}^{DC}:\underset{\mathbf{v}}{\mathrm{minimize}}\mathrm{\quad} & \tilde{p}\left(\mathbf{v}\left|\tilde{\omega}_{j},\omega_{jm}\right.\right)\label{eq:-11}\\
\textrm{subject to\quad} & \sum_{m=1}^{N_{G}}R_{m}\left(1-c_{q_{m},j}\right)\omega_{jm}\left\Vert \mathbf{v}_{jm}\right\Vert _{2}^{2}\leq C_{j}^{\mathrm{BH}},\forall j\in\mathcal{J}\tag{\ref{eq:-11}a}\nonumber \\
 & \left(\ref{eq:prob_p_original}\mathrm{b}\right),\left(\ref{eq:QoS-DC}\right),\nonumber 
\end{align}
where $\tilde{\omega}_{j}$ is a weight associated with the $j$-th BS, and $\omega_{jm}$ is a weight associated with the $j$-th BS and the $m$-th user group. Similar to \cite{Reweighted-L1_candes2008}, we develop the iterative weight update rules as
\begin{equation}
\tilde{\omega}_{j}=\frac{1}{\left\Vert \mathbf{\tilde{v}}_{j}\right\Vert _{2}^{2}+\tau},\ \mathrm{and}\ \omega_{jm}=\frac{1}{\left\Vert \mathbf{v}_{jm}\right\Vert _{2}^{2}+\tau},\forall j\in\mathcal{J},\forall m\in\mathcal{M},\label{eq:omega}
\end{equation}
with $\mathbf{\tilde{v}}_{j}$ and $\mathbf{v}_{jm}$ obtained from the previous iteration and a small constant parameter
$\tau>0$. Since the beamformer $\mathbf{\tilde{v}}_{j}$ (or $\mathbf{v}_{jm}$) with a lower transmit power usually has less impact, its transmit power should be encouraged to be further reduced, and eventually forced to zero, in order to switch off this BS (and its backhaul data delivery). Consequently, we are motivated to design weight updating rules (\ref{eq:omega}) where $\tilde{\omega}_{j}$ and $\omega_{jm}$ are inversely proportional to the transmit power. The small parameter $\tau>0$ is introduced to provide stability, and to ensure that a zero-valued component $\mathbf{\tilde{v}}_{j}$
(or $\mathbf{v}_{jm}$) does not strictly prohibit a nonzero estimate at the next step. Similar heuristic updating rules were also adopted in \cite{SparseBF_userCentricCluster}. 

It is observed that problem $\mathscr{P}^{DC}$ has a convex objective function, as well as DC constraints and convex constraints, and thus falls into the category of the general DC programming problems which take the following form: 
\begin{align}
\underset{\mathbf{x}}{\mathrm{minimize}}\mathrm{\quad} & f_{0}\left(\mathbf{x}\right)-h_{0}\left(\mathbf{x}\right)\label{eq:DC-general-form}\\
\textrm{subject to\quad} & f_{i}\left(\mathbf{x}\right)-h_{i}\left(\mathbf{x}\right)\leq0,i=1,\dots,m,\nonumber 
\end{align}
where $f_{i}\left(\cdot\right)$ and $h_{i}\left(\cdot\right)$, for
$i=0,\dots,m$, are convex functions. The concave-convex procedure (CCCP) \cite{CCCP} has been developed
to reach a local minimum of DC programming problems with a guaranteed convergence, where $\mathbf{x}_{t}$ can
be updated by solving the convex subproblem: 
\begin{align}
\underset{\mathbf{x}}{\mathrm{minimize}}\mathrm{\quad} & g_{0}\left(\mathbf{x}\left|\mathbf{x}_{t}\right.\right)\label{eq:CCCP-general-form}\\
\textrm{subject to\quad} & g_{i}\left(\mathbf{x}\left|\mathbf{x}_{t}\right.\right)\leq0,i=1,\dots,m,\nonumber 
\end{align}
where 
\begin{equation}
g_{i}\left(\mathbf{x}\left|\mathbf{x}_{t}\right.\right)=f_{i}\left(\mathbf{x}\right)-\left[h_{i}\left(\mathbf{x}_{t}\right)+\nabla h_{i}\left(\mathbf{x}_{t}\right)^{T}\left(\mathbf{x}-\mathbf{x}_{t}\right)\right],\label{eq:-1}
\end{equation}
for all $i=0,\dots,m$. To be specific, for problem $\mathscr{P}^{DC}$, the subproblem in the $t$-th iteration of the CCCP takes the following form: 
\begin{align}
\underset{\mathbf{v}}{\mathrm{minimize}}\mathrm{\quad} & \tilde{p}\left(\mathbf{v}\left|\tilde{\omega}_{j}^{\left[t\right]},\omega_{jm}^{\left[t\right]}\right.\right)\label{eq:-14}\\
\textrm{subject to\quad} & \begin{array}{cc}
\gamma_{m}\left(\sum_{i\neq m}^{N_{G}}\left|\mathbf{h}_{k}^{\mathsf{H}}\mathbf{v}_{i}\right|^{2}+\sigma_{k}^{2}\right)-2\mathrm{Re}\left\{ \left(\mathbf{v}_{m}^{\left[t\right]}\right)^{\mathsf{H}}\mathbf{h}_{k}\mathbf{h}_{k}^{\mathsf{H}}\mathbf{v}_{m}\right\} \\
+\left(\mathbf{v}_{m}^{\left[t\right]}\right)^{\mathsf{H}}\mathbf{h}_{k}\mathbf{h}_{k}^{\mathsf{H}}\mathbf{v}_{m}^{\left[t\right]}\leq0,\forall k\in\mathcal{G}_{m},\forall m\in\mathcal{M}
\end{array}\tag{\ref{eq:-14}a}\nonumber \\
 & \sum_{m=1}^{N_{G}}R_{m}\left(1-c_{q_{m},j}\right)\omega_{jm}^{\left[t\right]}\left\Vert \mathbf{v}_{jm}\right\Vert _{2}^{2}\leq C_{j}^{\mathrm{BH}},\forall j\in\mathcal{J}\tag{\ref{eq:-14}b}\nonumber \\
 & \sum_{m=1}^{N_{G}}\left\Vert \mathbf{v}_{jm}\right\Vert _{2}^{2}\leq P_{j}^{\mathrm{TX}},\forall j\in\mathcal{J},\tag{\ref{eq:-14}c}\nonumber 
\end{align}
where the coefficients
\begin{equation}
\tilde{\omega}_{j}^{\left[t\right]}=\frac{1}{\left\Vert \mathbf{\tilde{v}}_{j}^{\left[t\right]}\right\Vert _{2}^{2}+\tau},\ \mathrm{and}\ \omega_{jm}^{\left[t\right]}=\frac{1}{\left\Vert \mathbf{v}_{jm}^{\left[t\right]}\right\Vert _{2}^{2}+\tau}\label{eq:omega_iteration}
\end{equation}
are updated with the solution $\mathbf{v}^{\left[t\right]}$ obtained in the previous iteration. \textcolor{black}{In the $t$-th iteration, we obtain the solution $\mathbf{v}^{\left[t+1\right]}$ by solving problem (\ref{eq:-14}). Problem (\ref{eq:-14}) is a convex quadratically constrained quadratic program (QCQP), which can be regarded as a special case of a second-order cone program (SOCP) and can be readily solved by interior-point methods with complexity as $\mathcal{O}\left(N_{G}^{3.5}N_{B}^{3.5}L^{3.5}\right)$ \cite{ConvexOpt_book}.} To solve our problem efficiently, we need to carefully choose an initial feasible point for the CCCP algorithm. Therefore, an initialization step is proposed by solving a transmit power minimization problem $\mathscr{P}_{0}$ with SDR technique \cite{multicasting}, i.e., 
\begin{align}
\mathscr{P}_{0}:\underset{\left\{ \mathbf{W}_{m}\right\} }{\mathrm{minimize}}\mathrm{\quad} & \sum_{m=1}^{N_{G}}\mathrm{Tr}\left(\mathbf{W}_{m}\right)\label{eq:P_0}\\
\textrm{subject to\quad} & \frac{\mathrm{Tr}\left(\mathbf{W}_{m}\mathbf{H}_{k}\right)}{\sum_{i=1,i\neq m}^{N_{G}}\mathrm{Tr}\left(\mathbf{W}_{i}\mathbf{H}_{k}\right)+\sigma_{k}^{2}}\geq\gamma_{m},\forall k\in\mathcal{G}_{m},\forall m\in\mathcal{M}\tag{\ref{eq:P_0}a}\nonumber \\
 & \sum_{m=1}^{N_{G}}\mathrm{Tr}\left(\mathbf{W}_{m}\mathbf{J}_{j}\right)\leq P_{j},\forall j\in\mathcal{J}\tag{\ref{eq:P_0}b}\nonumber \\
 & \sum_{m=1}^{N_{G}}R_{m}\left(1-c_{q_{m},j}\right)\omega_{jm}\mathrm{Tr}\left(\mathbf{W}_{m}\mathbf{J}_{j}\right)\leq C_{j}^{\mathrm{BH}},\forall j\in\mathcal{J}\tag{\ref{eq:P_0}c}\nonumber \\
 & \mathbf{W}_{m}\succeq0,\forall m\in\mathcal{M},\tag{\ref{eq:P_0}d}\nonumber 
\end{align}
where we define two matrices $\mathbf{W}_{m}=\mathbf{v}_{m}\mathbf{v}_{m}^{H}\in\mathbb{C}^{N\times N},\forall m\in\mathcal{M}$ and $\mathbf{H}_{k}=\mathbf{h}_{k}\mathbf{h}_{k}^{H}\in\mathbb{C}^{N\times N},\forall k\in\mathcal{K}$,
to lift the quadratic constraints into higher dimensions. Moreover, we define a set of selective matrices $\mathbf{J}_{j}\in\left\{ 0,1\right\} ^{N\times N},\forall j\in\mathcal{J} $, with $\mathbf{J}_{j}=\mathrm{diag}\left(\mathbf{0}_{\left(j-1\right)L},\mathbf{1}_{L},\mathbf{0}_{\left(N_{B}-j\right)L}\right)$ as a diagonal matrix. All $\left\{ \mathbf{W}_{m}\right\} $ are rank-one constrained. If the solution $\left\{\mathbf{W}_{m}\right\} $ are all rank-one, the feasible beamformers $\left\{ \mathbf{v}_{m}\right\} $ obtained
by applying the eigenvalue decomposition (EVD) on $\left\{ \mathbf{W}_{m}\right\} $ can be directly employed as the initial feasible point for the CCCP algorithm. If the solutions $\left\{ \mathbf{W}_{m}\right\} $ are not rank-one,
$\left\{ \mathbf{v}_{m}\right\} $ are obtained through randomizing and scaling. If problem $\mathscr{P}_{0}$ is infeasible, the original problem $\mathscr{P}$ is infeasible and the optimization has to terminate. 

After solving problem $\mathscr{P}^{DC}$, we will obtain the sparse beamforming vector $\hat{\mathbf{v}}$ as the output of the first stage, as shown in Fig. \ref{fig:framework}. The algorithm solving the group sparsity-inducing norm minimization problem for the first stage is presented as Algorithm \ref{alg:stage I}, \textcolor{black}{which will converge to local minima or saddle points of problem $ \mathscr{P}^{DC} $ \cite{CCCP,lanckriet2009convergenceCCCP}, if it is feasible.}
\begin{algorithm}
\caption{The Group Sparsity-Inducing Norm Minimization Algorithm\label{alg:stage I} } \textbf{Step 1: }Find an initial feasible point $\left\{ \mathbf{v}^{\left[0\right]}\right\} $ by solving problem $\mathscr{P}_{0}$; 

\textbf{Step 2:} Initialize$\left\{ \omega_{jm}^{\left[0\right]}\right\} $,
$\left\{ \tilde{\omega}_{j}^{\left[0\right]}\right\} $, and set the iteration counter as $t=0$;

\textbf{Step 3:} Repeat
\begin{enumerate}
\item \textcolor{black}{Solve problem (\ref{eq:-14})}, and obtain the beamformer $\mathbf{v}^{\left[t+1\right]}$;
\item Set $t=t+1$, and update the weights according to (\ref{eq:omega_iteration});
\end{enumerate}
\textbf{Step 4: }Until stopping criterion is met\textbf{ }and obtain
the beamformer $\hat{\mathbf{v}}$;

\textbf{End}
\end{algorithm}

\subsection{Stage II: Iterative Search Procedure}

\begin{figure}
\centering{}\includegraphics[width=0.7\textwidth]{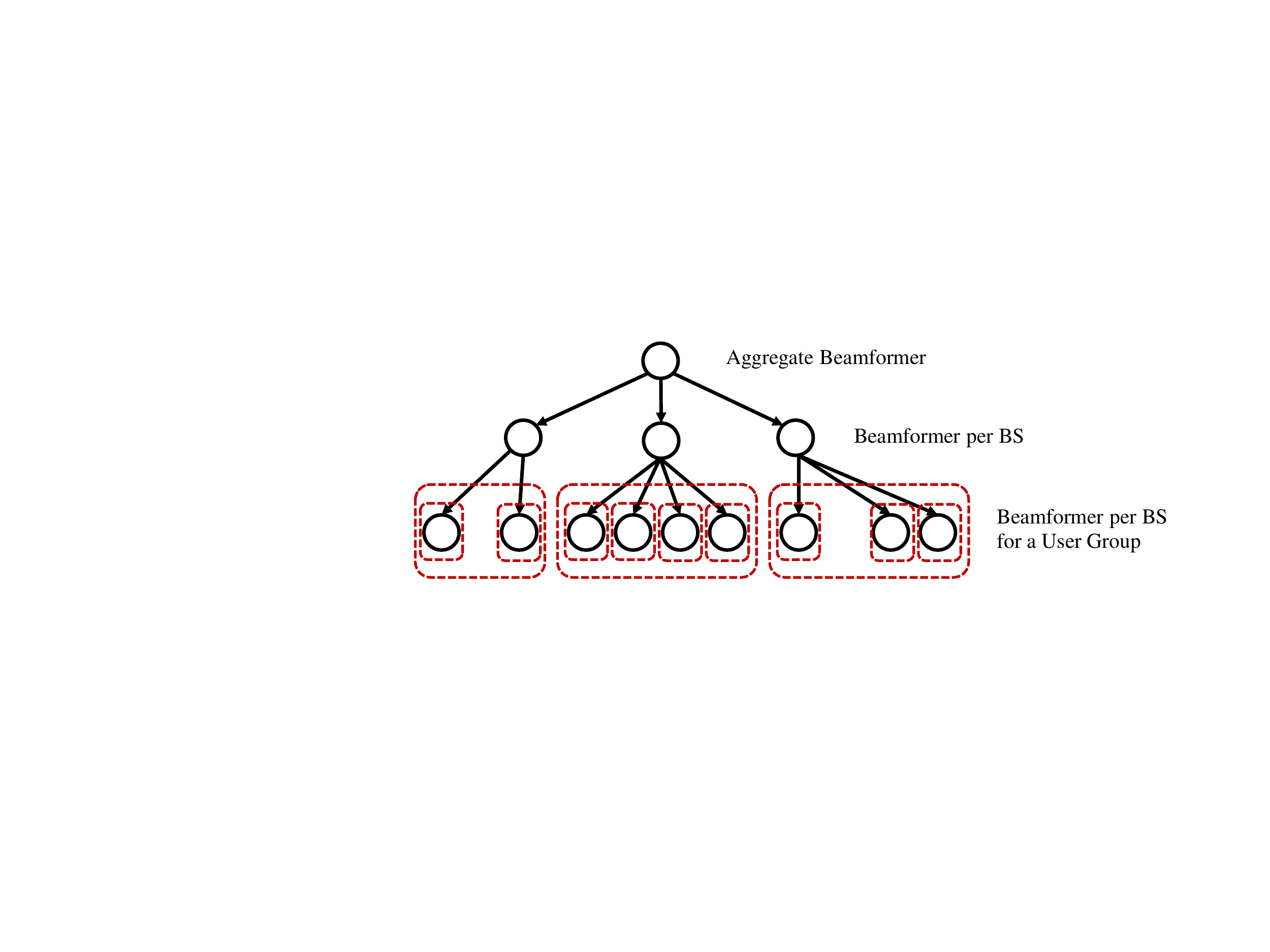}
\caption{\label{fig:The-layered-group}The layered group sparse pattern in the aggregate beamformer.}
\end{figure}

Inducing the sparsity structure in the solution is critical to problem
$\mathscr{P}^{LGSBF}$. As illustrated in Fig. \ref{fig:The-layered-group},
the layered group sparse pattern can be mapped to a hierarchy tree.
Generally, the solution $\hat{\mathbf{v}}$ obtained from Stage I
is not strictly sparse, and thus we propose to trim the entries of $\hat{\mathbf{v}}$
to obtain the group sparse solution. Although backhaul capacity constraints
can help us filter some sparsity patterns (i.e, prune some nodes in
the hierarchy tree), finding the optimal sparsity pattern in $\mathbf{v}$
brings about high computational complexity. In this subsection, we
develop an efficient search procedure to identify active BSs and backhaul
data assignment. 

\subsubsection{BS Ordering and Selection}

With the knowledge of the input $\mathbf{\hat{v}}$, the next step
is to determine the active BS set. After giving proper priorities
to BSs, we can obtain an ordering list to switch them off. Previous
works have considered different ways to calculate the priorities.
For example, Mehanna \emph{et al.} \cite{mehanna2013joint} directly
mapped the group-sparsity obtained by the group-sparsity inducing
norm minimization to their application, i.e., the transmit antennas
with smaller coefficients in the group were determined to be turned
off with a higher priority. Following this idea, in our setting, the
priorities might be given as $\tilde{\theta}_{j}=\left\Vert \mathbf{\tilde{v}}_{j}\right\Vert _{2},\forall j,$
which implies that the BS with a lower transmit beamforming gain should
be encouraged to be switched off. However, such a direct mapping might
bring performance degradation, as shown in \cite{GSBF}. To get a
better performance, it is essential to consider not only the transmit
beamforming gain but also other key system parameters indicating the
impact of the BSs on the network performance. Similar to the one employed
in \cite{GSBF}, to assign priorities to BSs, we propose the following
ordering criteria that incorporates channel power gain, BS power amplifier
efficiency, relative power consumption, backhaul power consumption,
caching status and beamforming gain, that is, 
\begin{equation}
\tilde{\theta}_{j}=\sqrt{\frac{\tilde{\kappa_{j}}}{\delta_{j}\left(P_{j}^{D}+\sum_{m=1}^{N_{G}}\beta_{jm}\left(1-c_{q_{m},j}\right)\right)}}\left\Vert \mathbf{\tilde{v}}_{j}\right\Vert _{2},\forall j,\label{eq:order-criterion1}
\end{equation}
where $\tilde{\kappa_{j}}=\sum_{k=1}^{N_{U}}\left\Vert h_{kj}\right\Vert ^{2}$
is the channel gain from the $j$-th BS to all MUs. The BS with a
higher priority (i.e., smaller $\tilde{\theta}_{j}$) will be switched
off before the one with a lower priority (i.e., larger $\tilde{\theta}_{j}$).
This ordering criteria implies that the BS with a lower channel power
gain, lower BS power amplifier efficiency (i.e., higher $\delta_{j}$),
higher relative transport link power consumption, higher backhaul
power consumption and lower cache hit ratio should have a higher priority
to be switched off.

Once BS $j$ is decided to be switched off, all its corresponding
beamforming coefficients will be set to zero, i.e, $\tilde{\mathbf{v}}_{j}=\mathbf{0}$.
Based on the ordering criteria rule, we sort the coefficients in ascending
order. Each time a BS is decided to be switched off, the inactive
BS set $\mathcal{Z}_{BS}$ will be updated and we check a feasibility
problem
\begin{align}
\mathscr{F}_{1}\left(\mathcal{Z}_{BS}\right):\mathrm{find}\mathrm{\quad} & \mathbf{v}\label{eq:-10}\\
\textrm{subject to\quad} & \sum_{m=1}^{N_{G}}\left\Vert \mathbf{v}_{jm}\right\Vert _{2}^{2}\leq P_{j}^{\mathrm{TX}},\forall j\notin\mathcal{Z}_{BS}\tag{\ref{eq:-10}a}\nonumber \\
 & \tilde{\mathbf{v}}_{j}=\mathbf{0},\textrm{if }j\in\mathcal{Z}_{BS}\tag{\ref{eq:-10}b}\nonumber \\
 & \left(\ref{eq:QoS-DC}\right),\left(\ref{eq:-11}\mathrm{a}\right),\nonumber 
\end{align}
which is a DC programming, and can be solved by the CCCP algorithm.

\subsubsection{Backhaul Data Assignment for Active BSs}
If problem $\mathscr{F}_{1}\left(\mathcal{Z}_{BS}\right)$ is feasible,
the next question is to determine the backhaul data assignment for
the active BSs in order to further reduce the power consumption. Similar
to the design idea for (\ref{eq:order-criterion1}), we calculate
priorities of backhaul data assignment for the active BSs by taking
the aforementioned key system parameters into consideration. Consequently,
we propose the following ordering criteria to determine which backhaul
data assignment should be turned off, i.e.,
\begin{equation}
\theta_{jm}=\begin{cases}
\sqrt{\frac{\kappa_{jm}}{\delta_{j}\left(P_{j}^{D}+\beta_{jm}\left(1-c_{q_{m},j}\right)\right)}}\left\Vert \mathbf{v}_{jm}\right\Vert _{2}, & \textrm{if }j\notin\mathcal{Z}_{BS}\\
0, & \textrm{if }j\in\mathcal{Z}_{BS}
\end{cases},\label{eq:order-criterion2}
\end{equation}
where $\kappa_{jm}=\sum_{k\in\mathcal{G}_{m}}\left\Vert h_{kj}\right\Vert ^{2}$
is the channel gain from the $j$-th BS to the MUs in the $m$-th
user group. Based on the ordering criteria, we delete a piece of data
assignment each time and update the inactive data assignment set $\mathcal{Z}_{DA}$.
With $\mathcal{Z}_{BS}$ and $\mathcal{Z}_{DA}$, the subproblem that
we need to solve takes the following form: 
\begin{align}
\mathscr{F}_{2}\left(\mathcal{Z}_{BS},\mathcal{Z}_{DA}\right):\mathrm{find}\mathrm{\quad} & \mathbf{v}\label{eq:-17}\\
\textrm{subject to\quad} & \mathbf{v}_{jm}=\mathbf{0},\forall\left(j,m\right)\in\mathcal{Z}_{DA}\nonumber \\
 & \left(\ref{eq:QoS-DC}\right),\left(\ref{eq:-11}\mathrm{a}\right),\left(\ref{eq:-10}\mathrm{a}\right),\left(\ref{eq:-10}\mathrm{b}\right),\nonumber 
\end{align}
which is also a DC program, and can be solved by the CCCP algorithm. 

Realizing that switching off as many BSs as possible may not result in a minimum total network power consumption, we are motivated to adopt a conservative strategy to determine the final active BS set and backhaul data assignment. To obtain the minimum network power
consumption, we iteratively search over all possible $\mathcal{Z}_{BS}$ and $\mathcal{Z}_{DA}$, and record the corresponding network power. By comparing all the recorded values, we can determine $\left(\mathcal{Z}_{BS}^{\star},\mathcal{Z}_{DA}^{\star}\right)$ that corresponds to the minimal network power consumption. Overall, the iterative search method can be accomplished via solving no more than $\frac{N_{G}N_{B}\left(N_{B}+1\right)}{2}$ DC problems.

\subsection{Stage III: Obtain Transmit Beamformers}
With the obtained inactive BS set $\mathcal{Z}_{BS}^{\star}$ and inactive data assignment set $\mathcal{Z}_{DA}^{\star}$, we can obtain the final beamforming vector by solving the following problem:
\begin{align}
\mathscr{P}^{Final}\left(\mathcal{Z}_{BS}^{\star},\mathcal{Z}_{DA}^{\star}\right):\underset{\mathbf{v}}{\mathrm{minimize}}\mathrm{\quad} & \sum_{j=1}^{N_{B}}\sum_{m=1}^{N_{G}}\delta_{j}\left\Vert \mathbf{v}_{jm}\right\Vert _{2}^{2}\label{eq:finalBF}\\
\textrm{subject to\quad} & \sum_{m=1}^{N_{G}}\left\Vert \mathbf{v}_{jm}\right\Vert _{2}^{2}\leq P_{j}^{\mathrm{TX}},\forall j\notin\mathcal{Z}_{BS}^{\star}\nonumber\\
 & \tilde{\mathbf{v}}_{j}=\mathbf{0},\forall j\in\mathcal{Z}_{BS}^{\star}\nonumber\\
 & \mathbf{v}_{jm}=\mathbf{0},\forall\left(j,m\right)\in\mathcal{Z}_{DA}^{\star}\nonumber\\
 & \left(\ref{eq:QoS-DC}\right),\left(\ref{eq:-11}\mathrm{a}\right)\nonumber,
\end{align}
\textcolor{black}{which is also a DC program. In principle, problem (\ref{eq:finalBF}) can be globally solved via the branch-and bound algorithm by extending the method developed in \cite{Lu2017_EfficientGlobalAlgo}. Such global optimization algorithms have high computational complexity, and cannot be applied in dense networks. Therefore, the CCCP algorithm is adopted to  efficiently obtain a local optimal solution.} The overall iterative LGSBF algorithm is summarized in Algorithm \ref{alg:iterative-search}. 
\begin{algorithm}
	\begin{centering}
	\caption{The Iterative LGSBF Algorithm\label{alg:iterative-search}}	
	\par\end{centering}
	
	\textbf{Step 1: }Solve problem $\mathscr{P}^{LGSBF}$ by applying Algorithm
	\ref{alg:stage I}: \textbf{if} it is infeasible, \textbf{go to End};
	otherwise, obtain $\hat{\mathbf{v}}$;
	
	\textbf{Step 2:} Calculate the ordering criterion (\ref{eq:order-criterion1}),
	and sort the values in the ascending order $\tilde{\theta}_{\pi_{1}}\leq\cdots\leq\tilde{\theta}_{\pi_{N_{B}}}$;
	
	\textbf{Step 3:} Initialize $\mathcal{Z}_{BS}^{\left[0\right]}=\emptyset$,
	and $i=0$;
	
	\textbf{Step 4:} Solve the optimization problem $\mathscr{F}_{1}\left(\mathcal{Z}_{BS}^{\left[i\right]}\right)$
	\begin{enumerate}
	\item \textbf{If} $\mathscr{F}_{1}\left(\mathcal{Z}_{BS}^{\left[i\right]}\right)$
	is feasible, 
	
	\begin{enumerate}
	\item Calculate the ordering criterion (\ref{eq:order-criterion2}), and
	sort the values in the ascending order $\tilde{\theta}_{\varpi_{1}}\leq\cdots\leq\tilde{\theta}_{\varpi_{N_{B}N_{G}}}$; 
	\item Initialize $\mathcal{Z}_{DA}^{\left[0\right]}=\emptyset$, and $k=0$;
	\item \textbf{Repeat} Solve the optimization problem $\mathscr{F}_{2}\left(\mathcal{Z}_{BS}^{\left[i\right]},\mathcal{Z}_{DA}^{\left[k\right]}\right)$,
	update the set $\mathcal{Z}_{DA}^{\left[k+1\right]}=\mathcal{Z}_{DA}^{\left[k\right]}\cup\left\{ \varpi_{k+1}\right\} $
	and $k=k+1$;\textbf{ }
	\item \textbf{Until} infeasible, obtain $\mathcal{S_{K}}^{\left[i\right]}=\left\{ 0,1,\dots,k-1\right\} $;
	\item Update the set $\mathcal{Z}_{BS}^{\left[i+1\right]}=\mathcal{Z}_{BS}^{\left[i\right]}\cup\left\{ \pi_{i+1}\right\} $
	and $i=i+1$, \textbf{go to} \textbf{Step 4};
	\end{enumerate}
	\item \textbf{If} $\mathscr{F}_{1}\left(\mathcal{Z}_{BS}^{\left[i\right]}\right)$
	is infeasible, obtain $\mathcal{S_{I}}=\left\{ 0,1,\dots,i-1\right\} $,
	\textbf{go to Step 5};
	\end{enumerate}
	\textbf{Step 5: }Obtain the optimal inactive BS set $\mathcal{Z}_{BS}^{\star}$
	and inactive data assignment set $\mathcal{Z}_{DA}^{\star}$ by solving
	$\left(\mathcal{Z}_{BS}^{\star},\mathcal{Z}_{DA}^{\star}\right)=\underset{i\in\mathcal{S_{I}},k\in\mathcal{S_{K}}^{\left[i\right]}}{\arg\min}p^{\star}\left(\mathcal{Z}_{BS}^{\left[i\right]},\mathcal{Z}_{DA}^{\left[k\right]}\right)$;
	
	\textbf{Step 6:} Obtain beamformers by solving problem $\mathscr{P}^{Final}\left(\mathcal{Z}_{BS}^{\star},\mathcal{Z}_{DA}^{\star}\right)$;
	
	\textbf{End}
\end{algorithm}
Note that the proposed approach provides a general framework for a multi-layer GSBF problem, where various group sparsity-inducing algorithms, e.g., the smoothed $\ell_{p}$-minimization \cite{smoothedLp}, can be applied in Stage I. 
\textcolor{black}{\subsection{Complexity and Convergence Analysis}
It has been shown that for the iterative search procedure, the number of general DC problems to be solved is no more than $\frac{N_{G}N_{B}\left(N_{B}+1\right)}{2}$. To obtain a local optimal solution for general DC programs, at each iteration of the CCCP-based algorithm, we need to solve a convex QCQP (or equivalently SOCP) problem with a complexity of $\mathcal{O}\left(N_{G}^{3.5}N_{B}^{3.5}L^{3.5}\right)$ by interior-point methods, which constitutes the main computational complexity of the proposed LGSBF algorithm. For large-scale networks, other approaches for solving large-sized SOCPs, e.g., the alternating direction method of multipliers (ADMM) method \cite{Boyd2011ADMM}, need to be explored. For unconstrained DC programs with differentiable objectives, it could converge superlinearly \cite{lanckriet2009convergenceCCCP}, while the convergence rate of general DC programs is still an open problem.}

\section{Simulation results}

\label{Sec:Simulation}

In this section, we simulate the performance of the proposed algorithm. We consider a hexagonal multicell network, where each BS is located at the center of a hexagonal cell whose radius is set to be $500\textrm{ }\mathrm{m}$, and MUs are uniformly and independently distributed in the network, excluding an inner circle of $50\textrm{ }\mathrm{m}$ around each BS. The channel between the $j$-th BS and the $k$-th user is modeled as $\mathbf{h}_{kj}=10^{-\left.L\left(d_{kj}\right)\right/20}\sqrt{\varphi_{kj}s_{kj}}\mathbf{g}_{kj},$
where $L\left(d_{kj}\right)$ is the path-loss at distance $d_{kj}$, $s_{kj}$ is the shadowing coefficient, $\varphi_{kj}$ is transmit antenna power gain and $\mathbf{g}_{kj}$ is the small scale fading coefficient. We adopt the standard cellular network parameters as presented in Table \ref{tab:Simulation-Parameters}.
\begin{table}
\centering \caption{Simulation Parameters} \label{tab:Simulation-Parameters}
\begin{tabular}{c|c}
\hline 
{\footnotesize{}{}Parameter}  & {\footnotesize{}{}Value}\tabularnewline
\hline 
{\footnotesize{}{}Transmit antenna power gain $\varphi_{kj}$}  & {\footnotesize{}{}$\unit[10]{dBi}$}\tabularnewline
{\footnotesize{}{}Path-loss at distance $d_{kj}$ (km)}  & {\footnotesize{}{}$148.1+37.6\log_{10}\left(d_{kj}\right)$}\tabularnewline
{\footnotesize{}{}Standard deviation of log-norm shadowing $\sigma_{s}$}  & {\footnotesize{}{}$\unit[8]{dB}$}\tabularnewline
{\footnotesize{}{}Small scale fading distribution $\mathbf{g}_{kj}$
}  & {\footnotesize{}{}$\mathcal{CN}\left(0,\mathbf{I}\right)$}\tabularnewline
{\footnotesize{}{}Noise power spectral density $\sigma_{k}^{2}$
}  & {\footnotesize{}{}$\unit[-172]{\left.dBm\right/Hz}$}\tabularnewline
{\footnotesize{}{}Bandwidth $B_{0}$}  & {\footnotesize{}{}$\unit[10]{\mbox{MHz}}$ }\tabularnewline
{\footnotesize{}{}Maximum BS transmit power $P_{j}^{\mathrm{TX}}$}  & {\footnotesize{}{}$\unit[1]{W}$}\tabularnewline
{\footnotesize{}{}Slope of the load-dependent power consumption $\delta_{j}$}  & {\footnotesize{}{}$4$}\tabularnewline
\hline 
\end{tabular}
\end{table}

The file library contains 100 pieces of content, whose popularity
follows a Zipf distribution with parameter $\gamma_{z}=1.2$. In \textcolor{black}{this popularity model}, a small $\gamma_{z}$ implies a flat popularity distribution, while
a large $\gamma_{z}$ means the opposite. BSs are assumed to have equal cache sizes. \textcolor{black}{We shall briefly show the role of cache by varying the cache size and caching strategies via simulations. Herein, we consider two widely-employed heuristic caching strategies, i.e., the most popular caching (MPC) \cite{videoAmazon} and probabilistic caching (ProbC) \cite{Tao2015arXiv,cacheSizeValue}. For MPC, each BS caches as many popular files as possible in accordance with the file popularity rank in the descending order. As for ProbC, each BS randomly caches files with the same probabilities as their request probabilities.} Assume that the SINR requirements for different user groups are the same,
i.e., $\gamma_{m}=\gamma,\forall m\in\left\{ 1,\dots,N_{G}\right\} $.

\subsection{Network Power Consumption}

Consider a network with $N_{B}=7$ BSs, each of which has two antennas,
and $N_{U}=15$ single-antenna MUs. We set the relative power consumption
as $P_{j}^{D}=\left[5.6+j-1\right]\mathrm{W},\forall j\in{\cal J}$,
backhaul energy coefficient as $E_{j}^{\mathrm{BH}}=\unit[1\times10^{-7}]{\left.J\right/bit}$,
and per-BS backhaul capacity as $\unit[500]{Mbps}$. Each BS has a
cache size of 10 files \cite{cacheSizeValue}, i.e., $c_{fj}=1,\forall f=1,\dots,10,\forall j\in{\cal J}$. 

\begin{figure}
	\begin{centering}	 
	\subfigure[\scriptsize With MPC strategy.]{\centering{}\includegraphics[width=0.4\textwidth]{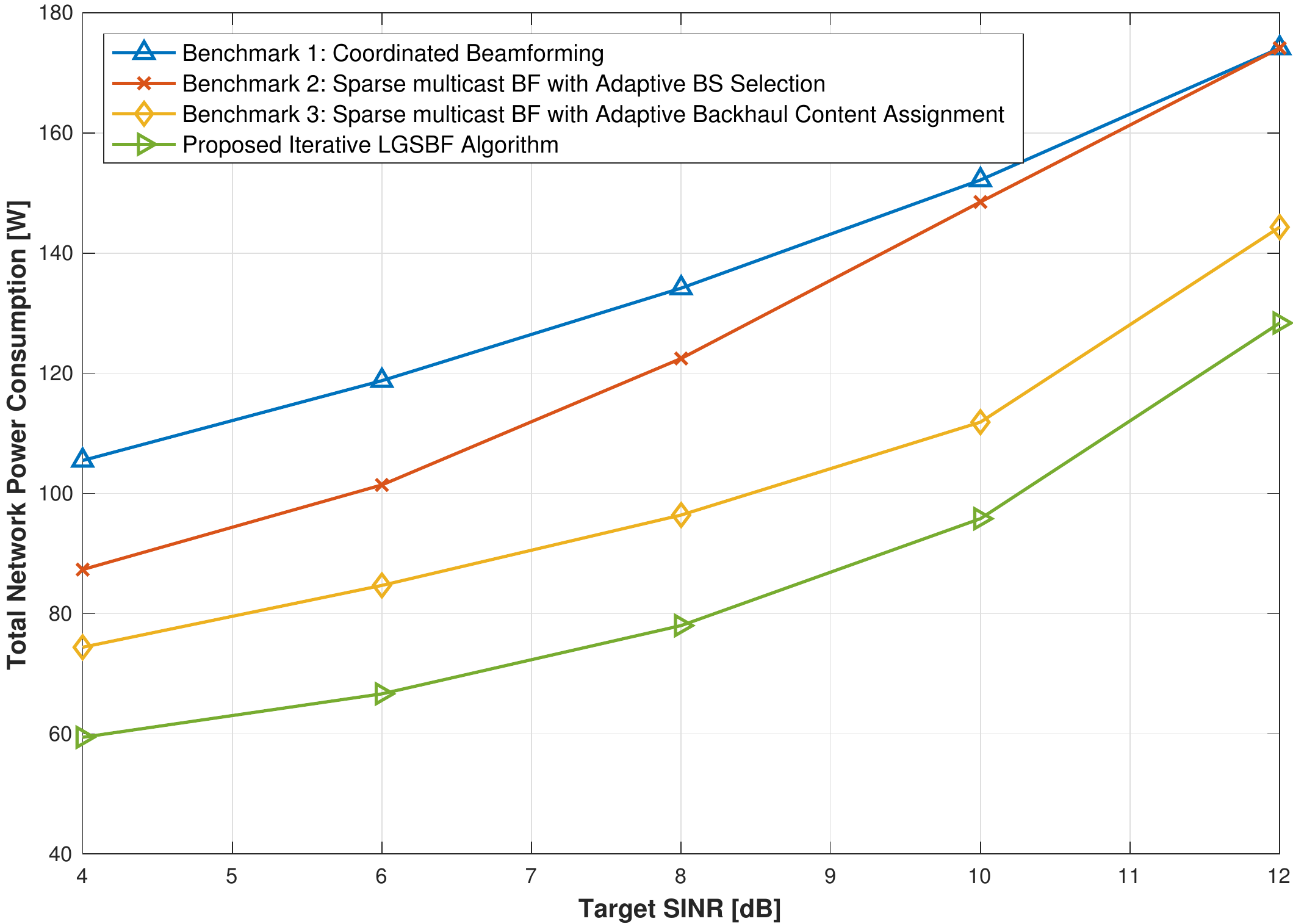}\label{fig: MPC} } $\qquad$	
	\subfigure[\scriptsize With ProbC strategy.]{\centering{}\includegraphics[width=0.40\textwidth]{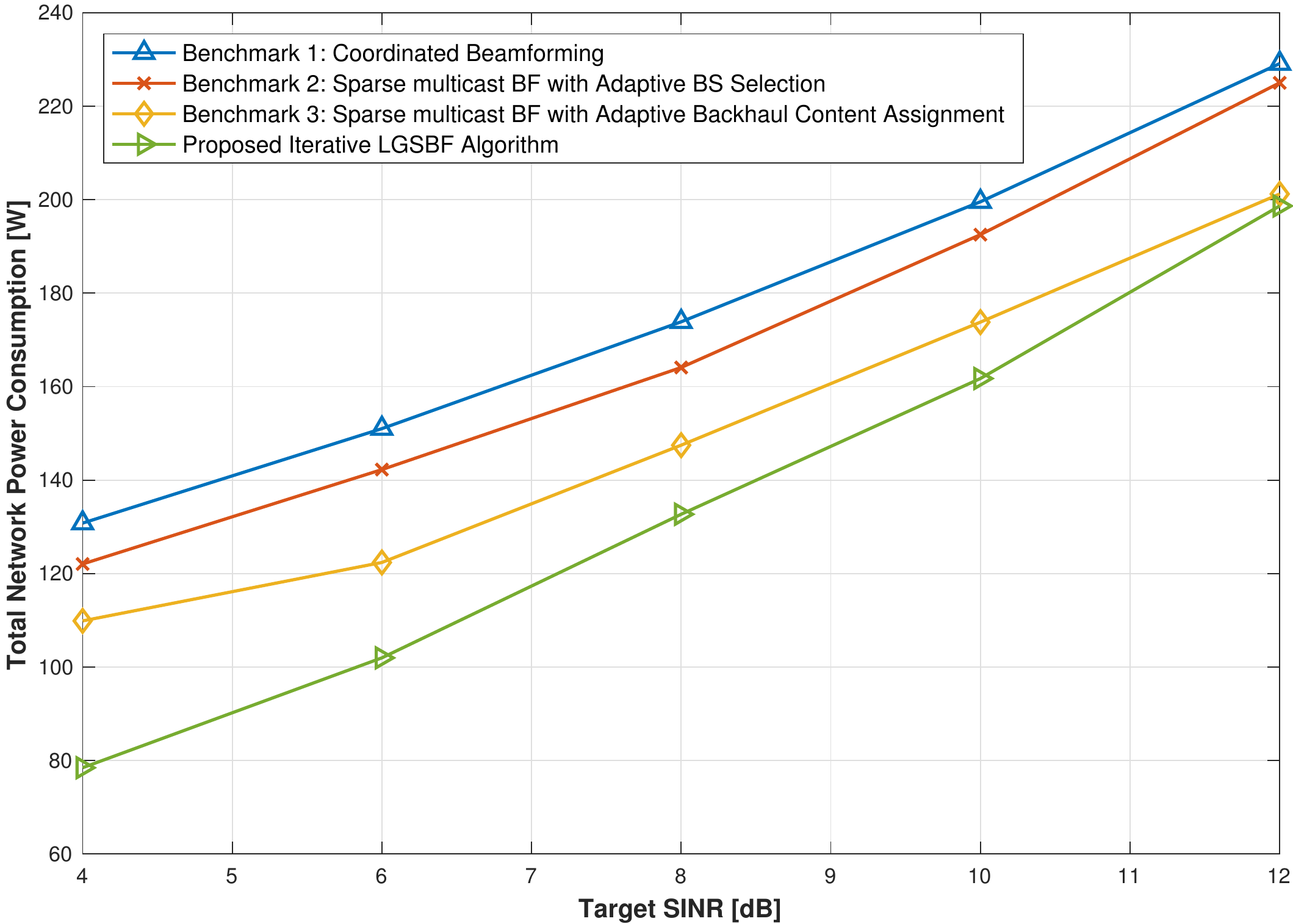}\label{fig: ProbC}}
	\par\end{centering}
\caption{\label{fig:Network-power-consumption}Network power consumption versus target SINR. }
\end{figure}

The proposed algorithm is compared with the following algorithms: 
\begin{itemize}
\item Coordinated beamforming (CB) algorithm: In this algorithm \cite{CB}, all BSs
are in the active mode and only the total BS transmit power consumption
is minimized. 
\item Sparse multicast beamforming algorithm with adaptive BS selection: This algorithm \cite{smoothedLp} develops a procedure to switch off as many BSs as possible. The non-convex smoothed $\ell_{p}$-norm is adopted to replace
the convex mixed $\left.\ell_{1}\right/\ell_{2}$-norm in the objective function. The non-convex quadratic forms of beamforming vectors in the objective function and the non-convex quadratic QoS constraints are relaxed by leveraging SDR technique. Then an iterative reweighted-$\ell_{2}$ algorithm is employed to solve the problem. 
\item Sparse multicast beamforming algorithm with adaptive backhaul content assignment: In this algorithm \cite{Tao2015arXiv}, the number of backhaul content assignments is minimized. Smooth approximated functions are employed to approximate the $\ell_{0}$-norm terms and a generalized CCCP algorithm is applied to solve the problem. \textcolor{black}{The arctangent function is adopted since \cite{Tao2015arXiv} shows it gives the best approximation performance.}
\end{itemize}
Fig. \ref{fig:Network-power-consumption} demonstrates the total network power consumption with different target SINR values \textcolor{black}{under MPC and ProbC, respectively}. It shows that the proposed iterative LGSBF algorithm outperforms existing algorithms \textcolor{black}{with both caching strategies}, which confirms the effectiveness of the proposed algorithm. \textcolor{black}{When the target SINR increases, it is observed that the gap between different algorithms becomes smaller, while benchmark 2 converges to CB faster than benchmark 3 and the proposed algorithm. It is because that more and more BSs need to be switched on to support the increasing QoS requirements, which decreases the benefit of active BS selection. Whereas, a careful design for backhaul content assignment can still help reduce network power consumption, since it can bring some cooperation chance for BSs and avoid unnecessary backhaul consumption at the same time. }

\textcolor{black}{
	\textbf{Remark 1}. The proposed algorithm achieves a better performance than those of \cite{smoothedLp} and \cite{Tao2015arXiv} by considering a two-layer adaptive selection for both active BS and actual backhaul assignment instead of the existing one-layer approaches. 
	This indicates that the joint adaptive decision of BS selection and backhaul content assignment can effectively reduce network power consumption for a wide range of target SINRs.}

\textcolor{black}{	
	\textbf{Remark 2}. Comparing two caching strategies, we observe that MPC performs better than ProbC in reducing network power consumption, and the gap becomes larger when the target SINR increases. In general, MPC provides better performance than ProbC for normal network settings, and similar findings are also observed in \cite{Tao2015arXiv}.
 }
%

\subsection{Impact of Cache Size}

\begin{figure}
\begin{centering}
\subfigure[\scriptsize Target SINR = 5 dB.]{\centering\includegraphics[width=0.45\textwidth]{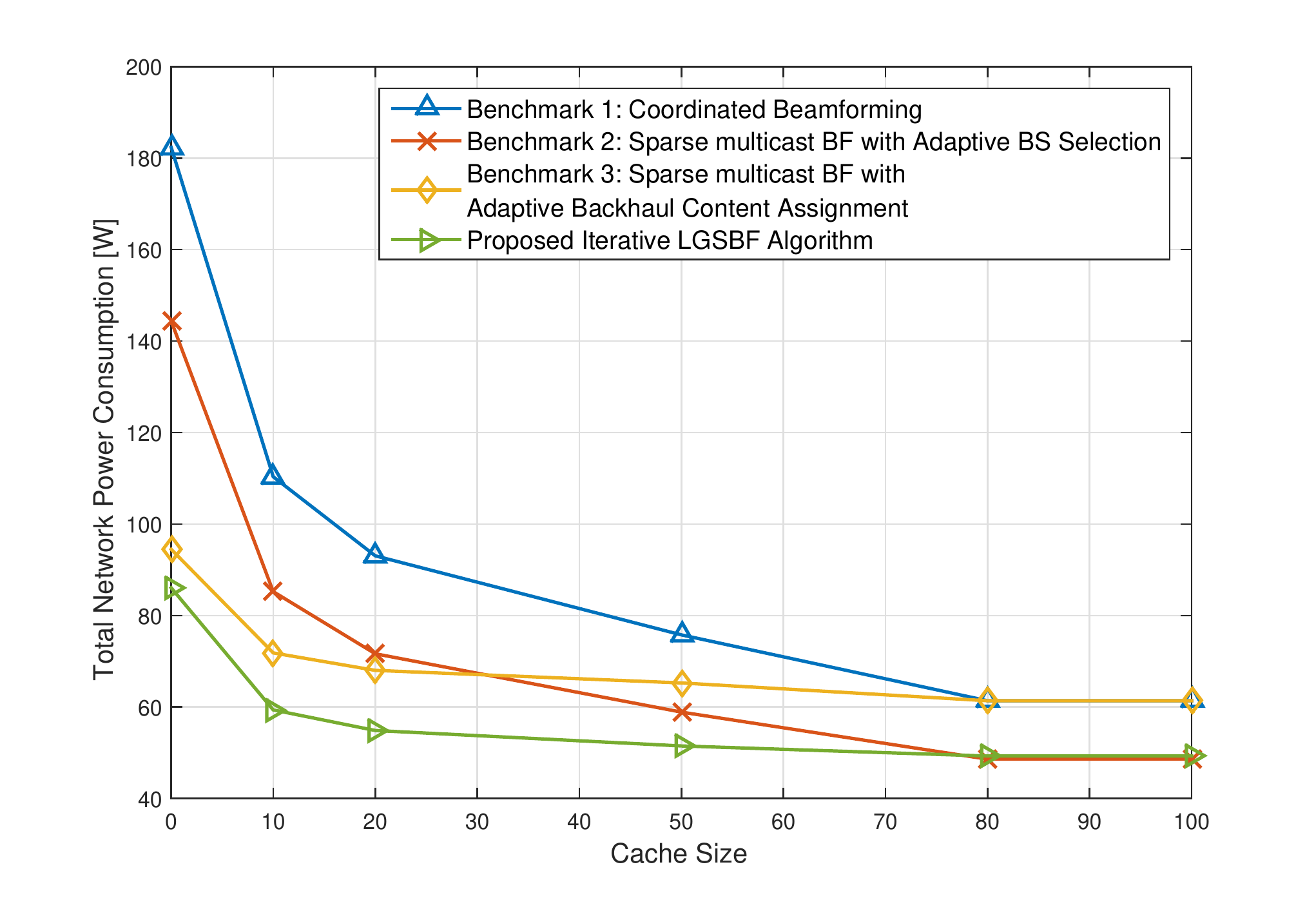} 
\label{fig:numBF-csize-5dB}}
\subfigure[\scriptsize Target SINR = 10 dB.]{\centering{}\includegraphics[width=0.45\textwidth]{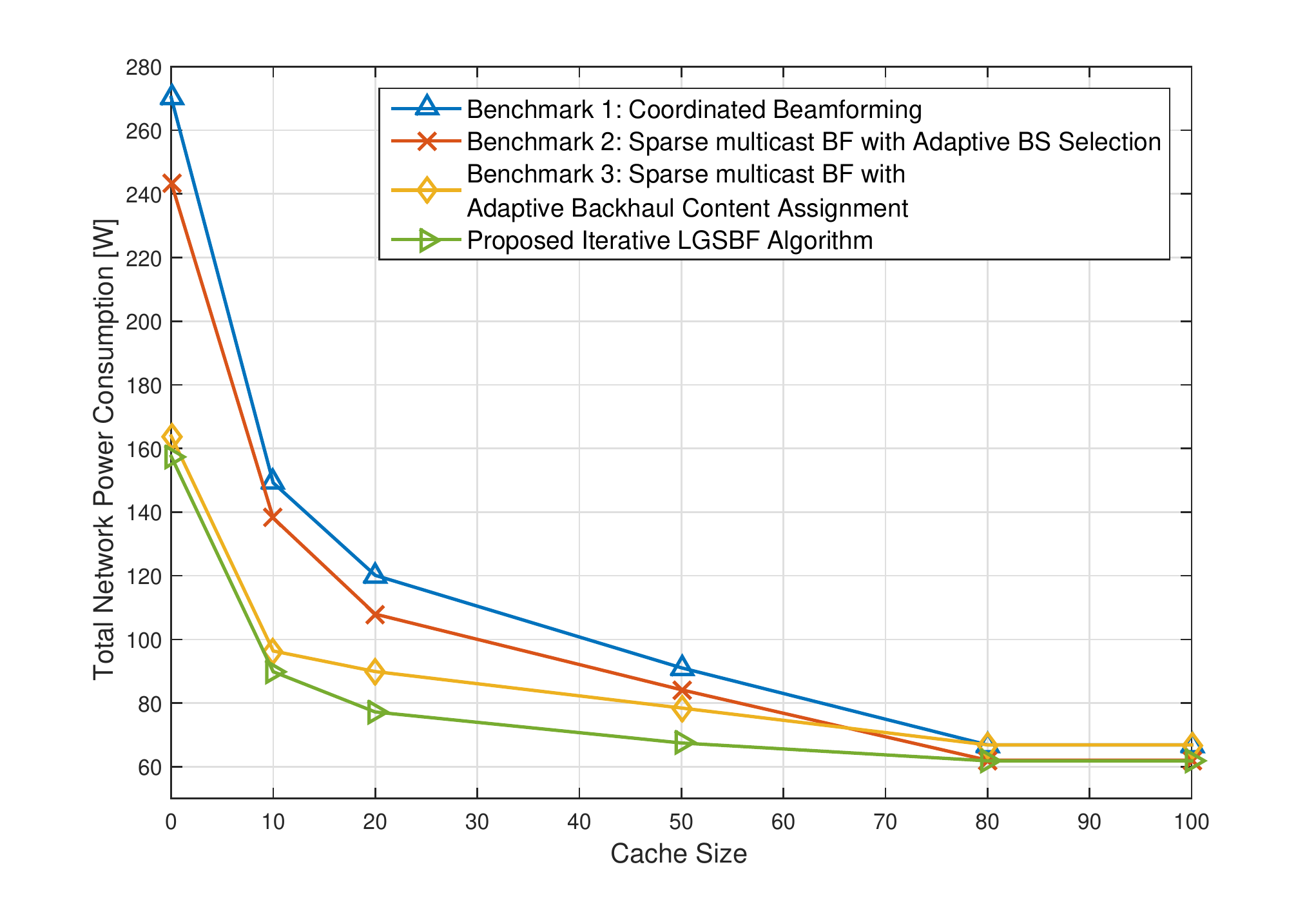}
\label{fig:numBF-csize-10dB}}
\par\end{centering}

\caption{\label{fig:Different-SINR}The tradeoff between the total network
power consumption and the cache size of each BS. }
\end{figure}

In Fig. \ref{fig:numBF-csize-5dB} and Fig. \ref{fig:numBF-csize-10dB},
we compare the performance of the proposed algorithm with benchmarks
in terms of the tradeoff between total network power consumption and
per-BS cache size under target $\textrm{SINR}=\unit[5]{dB}$ and target
$\textrm{SINR}=\unit[10]{dB}$, respectively.
Other settings are the same as those in Fig. \ref{fig:Network-power-consumption}.
From Fig. \ref{fig:Different-SINR}, it is observed that the proposed algorithm
outperforms benchmarks under different cache sizes. Moreover, the advantage of adaptive backhaul
content assignment will gradually be surpassed by adaptive BS selection when the cache size increases;
at a higher target SINR regime, the crosspoint will occur at a larger
cache size. Besides this, the proposed algorithm achieves a better
performance in a low SINR regime. It can be inferred that the increase
in the cache size allows more BSs to be switched off, especially when
the QoS requirement is comparatively low. 

\begin{figure}
\begin{centering}
\subfigure[\scriptsize The number of active BSs versus cache size.]{\centering{}\includegraphics[width=0.45\textwidth]{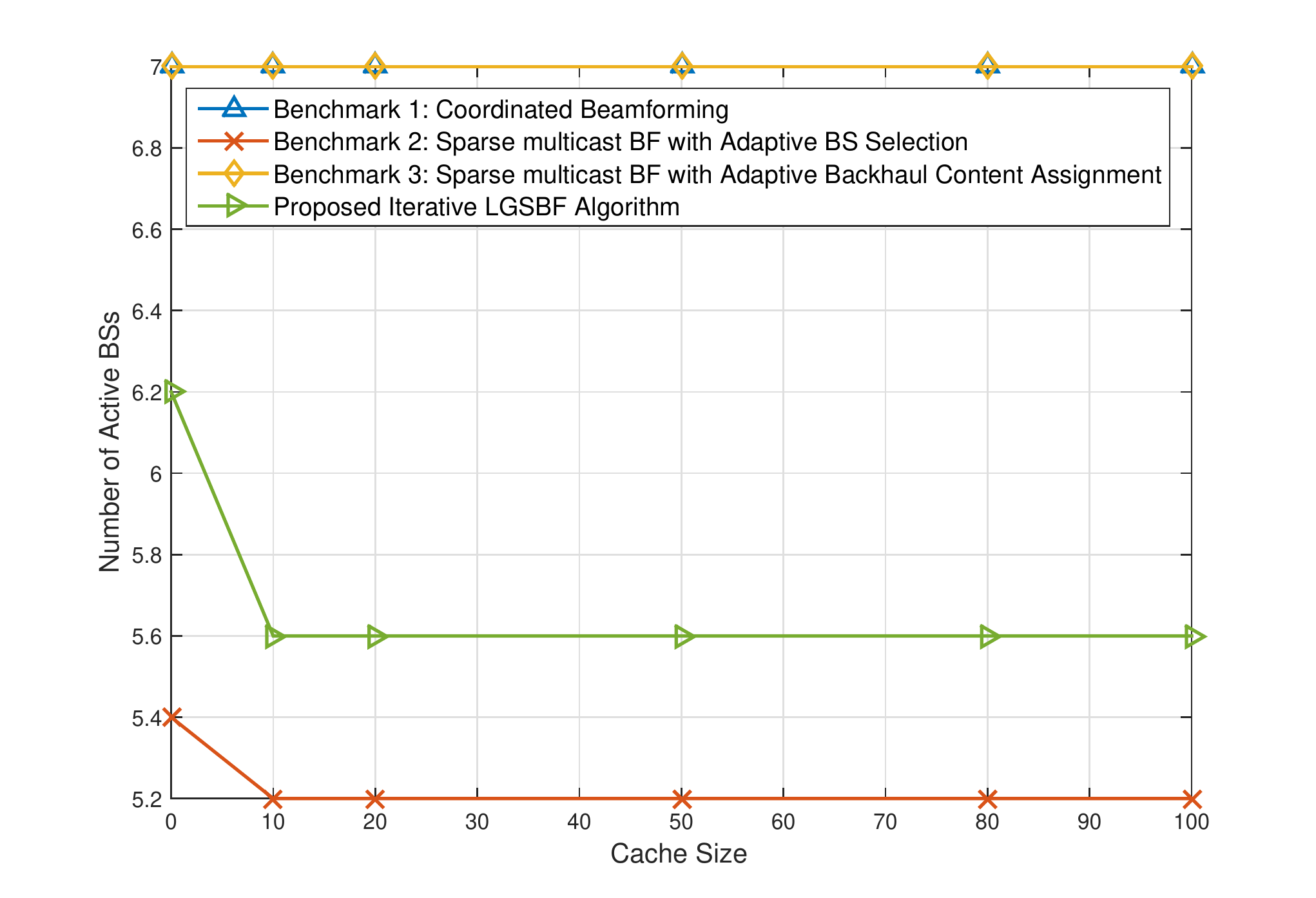} 
\label{fig: num-of-users}}
\subfigure[\scriptsize Backhaul power consumption versus cache size.]{\centering{}\includegraphics[width=0.45\textwidth]{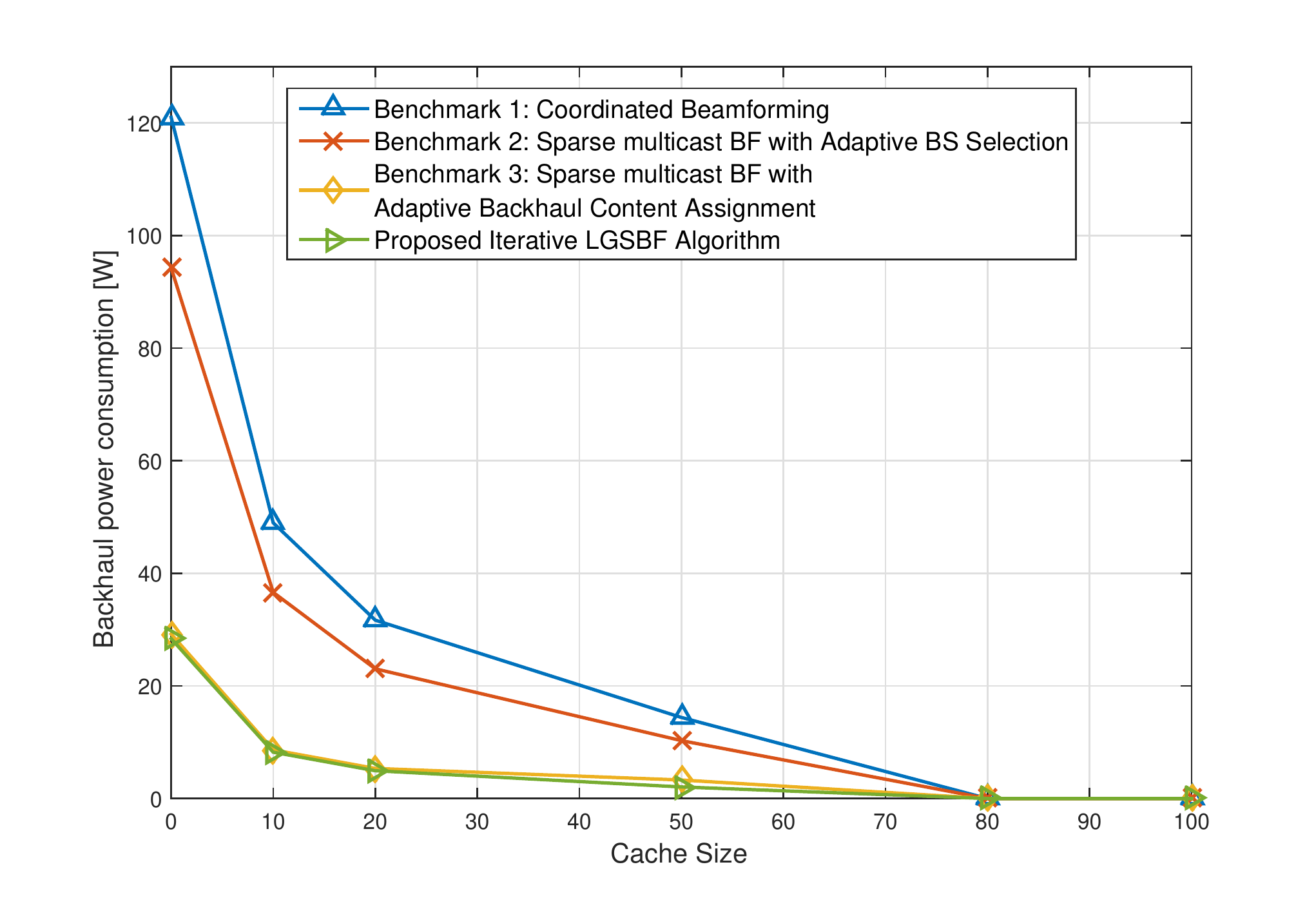}

\label{fig: backhaul-power}}
\par\end{centering}
\caption{\label{fig:Impact-of-cache-size}The impact of the cache size.}
\end{figure}

With the same network setting as in Fig. \ref{fig:numBF-csize-5dB},
the details of the impact of caching on the BSs and backhaul links
are demonstrated in Fig. \ref{fig:Impact-of-cache-size}. This figure shows that the CB algorithm, which intends to
minimize the BS transmit power consumption, has the highest backhaul
power consumption. This is because all the BSs are active in the CB
algorithm in order to achieve the highest beamforming gain. Moreover,
by comparing Benchmark 2, Benchmark 3 and the proposed algorithm,
it can be inferred that minimizing either the number of active BSs
or the number of backhaul content delivery cannot be the optimal strategy
to save power. Since both the backhaul power consumption and BS power
consumption hold a nontrivial share, a joint adaptive BS selection,
backhaul data assignment and power minimization beamforming is crucial
for minimizing the total network power consumption.

\subsection{Impact of the Number of Mobile Users and Backhaul Energy Coefficient}

\begin{figure}
\centering{}\includegraphics[width=0.5\textwidth]{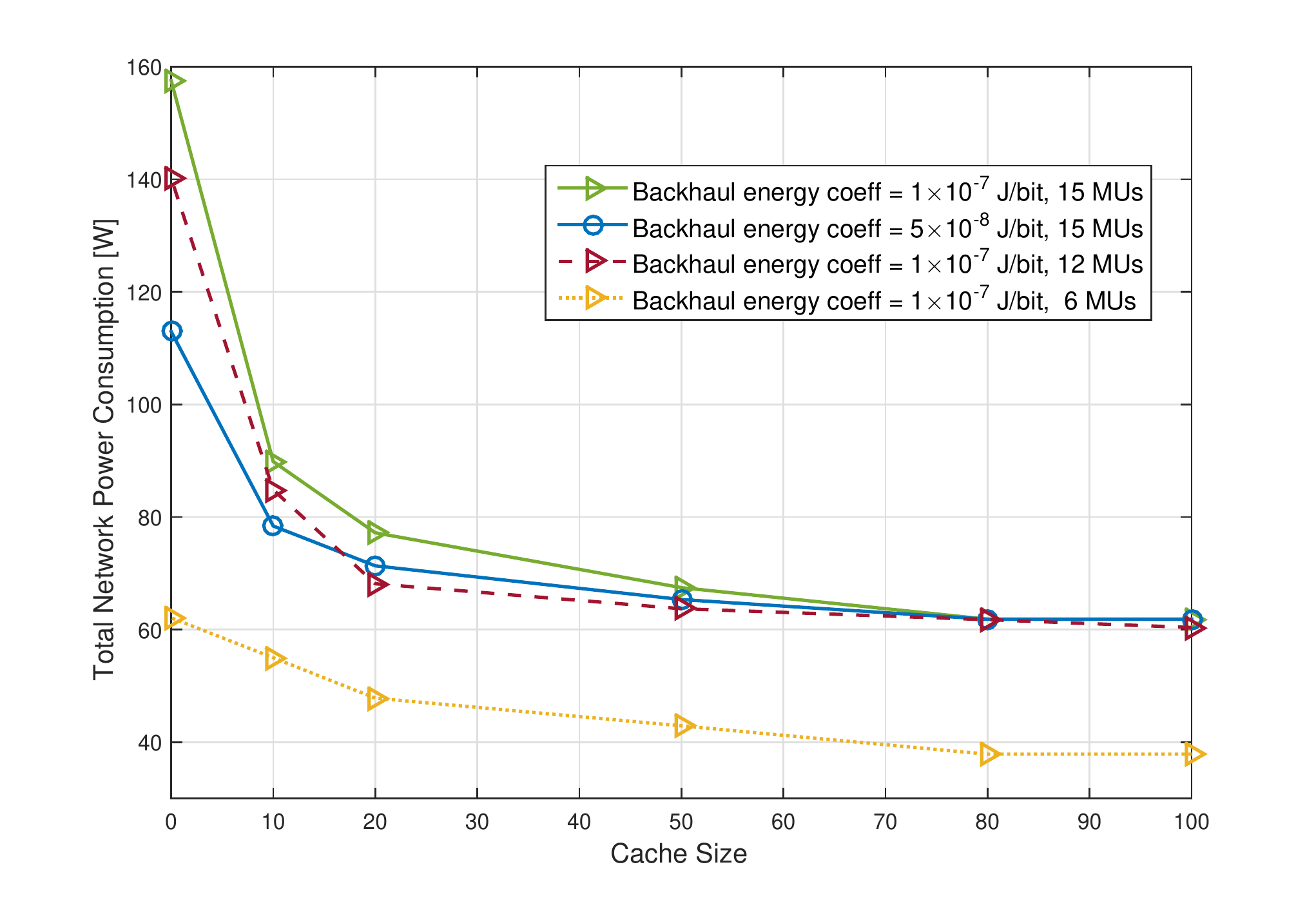} 
\caption{Network power consumption versus cache size under different MU densities.}
\label{fig:Network-power-Mu-num}
\end{figure}

We also investigate the impact of other important network parameters,
i.e., the number of MUs and backhaul energy coefficient, as shown
in Fig. \ref{fig:Network-power-Mu-num}. The figure demonstrates
that when the number of MUs increases, the performance gap between
the zero-cache case and full-cache case becomes larger. On the other
hand, Fig. \ref{fig:Network-power-Mu-num} also shows that the performance
gap between the zero-cache case and full-cache case is larger for
the network with a higher backhaul energy coefficient. Actually, different
backhaul energy coefficients represent different types of backhaul
links: a higher backhaul energy coefficient stands for less power-efficient
backhaul links, and vice versa. To enhance the performance in total
network power consumption, the operators can either upgrade the backhaul
links, which is expensive, or simply install cost-effective caches.
From the simulation, we can infer that caches will play a
more significant part in networks with higher user densities, and less
power-efficient backhaul links.

\section{Conclusions} \label{Sec:Conclusions}

In this study, we developed an effective framework to minimize the total network power consumption of cache-enabled wireless networks. The proposed LGSBF formulation generalized existing works on group sparse beamforming, for which an effective algorithm was developed. The proposed algorithm can significantly reduce the total network power consumption via a joint design of adaptive BS selection, backhaul content assignment and multicast beamforming. From the simulations, the proposed LGSBF framework was demonstrated to outperform existing algorithms by striking a balance between the BS power consumption and backhaul power consumption. Furthermore, it was shown that caching tends to play a more significant part in networks with higher user densities and less power-efficient backhaul links. For future research directions, it would be interesting to optimize the caching placement in the prefetching phase, and incorporate it when minimizing the total network power consumption. It is also important but challenging to develop more efficient distributed algorithms for practical implementation in large-scale networks. 


\bibliographystyle{IEEEtran}
\bibliography{IEEEabrv,Cachereport}

\end{document}